\documentclass{jfm}
\pdfoutput=1
\usepackage[english]{babel}
\usepackage[T1]{fontenc}
\usepackage[utf8]{inputenc}
\usepackage{amsmath,amssymb}
\usepackage[svgnames]{xcolor}
\usepackage{graphicx}
\usepackage{acro}
\usepackage{subfig}
\usepackage{hyperref}

\usepackage[normalem]{ulem}
\usepackage{soul}

\graphicspath{{./figures/}}

\DeclareAcronym{pdf}{
	short=PDF,
	long=Probability Density Function,
}
\DeclareAcronym{iid}{
	short=i.i.d.,
	long=Independent Identically Distributed
}
\DeclareAcronym{ou}{
	short=OU,
	long=Ornstein-Uhlenbeck,
}
\DeclareAcronym{gktl}{
	short=GKTL,
	long=Giardina-Kurchan-Tailleur-Lecomte,
}
\DeclareAcronym{ams}{
	short=AMS,
	long=Adaptive Multilevel Splitting,
}
\DeclareAcronym{tams}{
	short=TAMS,
	long=Trajectory Adaptive Multilevel Splitting,
}
\DeclareAcronym{scgf}{
	short=SCGF,
	long=Scaled Cumulant Generating Function,
}
\DeclareAcronym{lbm}{
	short=LBM,
	long=Lattice Boltzmann Method,
}
\DeclareAcronym{lbe}{
	short=LBE,
	long=Lattice Boltzmann Equation,
}
\DeclareAcronym{lgca}{
	short=LGCA,
	long=Lattice Gas Cellular Automata,
}
\DeclareAcronym{lbgk}{
	short=LBGK,
	long=Lattice Bhatnagar-Gross-Krook
}
\DeclareAcronym{OU}{
	short=O-U,
	long=Ornstein--Ulhenbeck
}
\DeclareAcronym{dns}{
	short=DNS,
	long=Direct Numerical Simulation,
}
\DeclareAcronym{md}{
	short=MD,
	long=Molecular Dynamics,
}
\DeclareAcronym{cfd}{
	short=CFD,
	long=Computational Fluid Dynamics,
}
\DeclareAcronym{pde}{
	short=PDE,
	long=Partial Differential Equation,
}

\title{
Numerical study of extreme mechanical force exerted by a turbulent flow on a bluff body by direct and rare-event sampling techniques
}

\author{Thibault Lestang\aff{1}\aff{2}
  \corresp{\email{thibault.lestang@cs.ox.ac.uk}},
  Freddy Bouchet\aff{1}
  \and Emmanuel L\'evêque\aff{2}}
\affiliation{\aff{1}Univ Lyon, ENS de Lyon, Univ Claude Bernard de Lyon, CNRS, Laboratoire de Physique, F-69342 Lyon, France
\aff{2}Univ Lyon, Ecole Centrale de Lyon, Univ Claude Bernard de Lyon, INSA de Lyon, CNRS, Laboratoire de M\'ecanique des Fluides et d'Acoustique, F-69134 Ecully cedex, France}
\begin{document}
	
\maketitle

\begin{abstract}
	This study investigates, by means of numerical simulations, extreme mechanical force exerted by a turbulent flow impinging on a bluff body, and examines the relevance of two distinct rare-event algorithms to efficiently sample these events. 
	The drag experienced by a square obstacle placed in a turbulent channel flow (in two dimensions) is taken as a representative case study.
	Direct sampling shows that extreme fluctuations are closely related to the presence of a strong vortex blocked in the near wake of the obstacle. This vortex is responsible for a significant pressure drop between the forebody and the base of the obstacle, thus yielding a very high value of the drag. 
	Two algorithms are then considered to speed up the sampling of such flow scenarii, namely the \ac{ams} and the \ac{gktl} algorithms. 
	The general idea behind these algorithms is to replace a long simulation by a set of much shorter ones, running in parallel, with dynamics that are replicated or pruned, according to some specific rules designed to sample large-amplitude events more frequently.  These algorithms have been shown to be relevant for a wide range of problems in statistical physics, computer science, biochemistry.
        The present study is the first application to a fluid-structure interaction problem. 
	% allowing the simulation of rare  events otherwise out of reach by direct sampling. 
	%
	Practical evidence is given that the fast sweeping time of turbulent fluid structures past the obstacle has a strong influence on the efficiency of the rare-event algorithm. 
	While the \ac{ams} algorithm does not yield significant run-time savings as compared to direct sampling, the \ac{gktl} algorithm appears to be effective to sample very efficiently extreme fluctuations of the time-averaged drag and estimate related statistics such as return times.
      Software used for simulations and data processing is available at \url{https://github.com/tlestang/paper_extreme_drag_fluctuations}.
\end{abstract}
	
\section{Introduction}

% general comments on physical problem %
%
Turbulent flows are important in a variety of natural phenomena, industrial and civil applications.
Their characteristic feature is the spontaneous development of intense and sporadic motions associated with extreme internal forces \citep{lesieur_book,donzis_sreenivasan_2010,Yeung}.
``Extreme'' refers here to fluctuations that can deviate from the mean value by ${\cal{O}}(10)$ standard deviations.
In engineering, the nature of such extreme dynamical events and their statistics are of crucial interest to predict excessive force, which can threaten the structural integrity of embedded structures \citep{kanev2010}.

From the viewpoint of chaotic dynamical systems, turbulence in fluids is linked to non-linearity and strong departure from statistical equilibrium \citep{KRAICHNAN}.
The use of analytical perturbative methods in identifying resonant interactions (among degrees of freedom) responsible for extreme fluctuations is unsuccessful.
Alternatively, simulation offers a practical approach to gain physical insight into these events, quantifying their intensity and estimating their frequency of occurrence.
However, this requires very long simulations since these extreme events are rare. The computational cost for sampling a fluctuation of very small probability typically grows as the inverse of this probability~\citep{wouters2016rare}.
{Rare-event sampling refers to a large body of methods that aim at \emph{preferentially} exploring  regions of phase space corresponding to  events that would otherwise be accessed with a very low probability through a brute-force direct sampling.}
%
%	
% next two paragraphs may be too technical -- Find a way to summarize them in a few sentences accessible to a broad audience in fluid mechanics? Notion of action will be vague for many readers
%
In the present work, a computational study of extreme mechanical force acting on an immersed bluff body is conducted using both very long time-series (direct sampling) and rare-event sampling techniques.

In fluid turbulence, rare-event sampling has been approached mainly from the perspective of simplified dynamics such as the one-dimensional Burgers' equation with a stochastic forcing \citep{bec_burgers_2007}. In this case, dynamics can be sampled by using a Markov chain Monte-Carlo algorithm \citep{duben_monte_2008,mesterhazy2011anomalous,mesterhazy2013lattice} that provides a framework for rare-event sampling.
An alternative approach is based on instantons \citep{gurarie_instantons_1996,grafke2015instanton} and applies to stochastically driven systems in the limit of weak noise.
Instantons refer to the most probable trajectories in phase space that achieve a given rare event (in the limit of weak noise). Suitable numerical schemes can be used to evaluate instantons as well as the related probabilities of rare events \citep{chernykh_large_2001,grafke_instanton_2013,grigorio_instantons_2017,laurie2015computation,bouchet2014langevin}.
An example is the investigation of the physics of rogue waves~\citep{dematteis2018rogue,dematteis2019experimental}.
However, a drawback of the aforementioned approaches is their limitation to simple and stochastically driven dynamics.
Here, a more general approach is considered for complex, possibly deterministic, dynamical systems.
It is based on sampling algorithms relying on \emph{selection rules} applied to an ensemble of trajectories of the system. 
Even though such ideas date back to the early 1950s, they have received growing interest over the last twenty years with successful applications in various domains such as chemistry \citep{van_erp_elaborating_2005,escobedo_transition_2009,teo_adaptive_2016}, biophysics \citep{huber_weighted-ensemble_1996,zuckerman2017weighted,bolhuis2005kinetic}, nuclear physics \citep{louvin2017}, nonlinear dynamical systems \citep{tailleur_probing_2007} and communication networks simulation \citep{villen-altamirano_restart:_1994}.
More importantly, such algorithms have been shown to be useful for the study of rare events in simple deterministic dynamics~\citep{wouters2016rare}.
Certainly, an original contribution of our study is to test the application of rare-event sampling algorithms in the context of far-from-equilibrium dynamics with an irreducible very large number of degrees of freedom. 
Two algorithms that are a priori suitable for such dynamics are considered, namely, the Adaptive Multilevel Splitting (AMS) algorithm \citep{cerou_adaptive_2007} and the Giardina-Kurchan-Tailleur-Lecomte (GKTL) algorithm \citep{giardina_direct_2006}.
Another problem related to rare events is the estimation of parameters or statistics from a limited number of samples.  Several authors have proposed original strategies  \citep{Mohamad-ws,Blonigan-ws} such as sequential sampling. These approaches and rare event algorithms could also be considered for applications in wave-structure problems, ship motion and load acting on offshore platforms \citep{belenky-ws}.
The paper is organized in two parts. The first part highlights the phenomenology of extreme fluctuations of the drag force acting on a square placed in a two-dimensional turbulent channel flow. This study is based on the simulation of the flow over a very long duration, made possible by the relative simplicity of the flow.
The motivation for this study is twofold.
Firstly, it provides a detailed description of the statistics and dynamics related to extreme drag fluctuations. This analysis is informative from the  viewpoint of fluid mechanics and, to the best of our knowledge, has never been reported before.
Secondly, it yields reference results that are required to validate the outputs of rare-event algorithms and to evaluate the possible computational gain obtained from them. This assessment is developed in the second part of the paper.

% Annonce du plan
The flow set-up is introduced and the dynamics related to typical
drag fluctuations is described in section~\ref{sec:test_flow}.
The statistical properties of the drag are then discussed.
In section~\ref{sec:direct_sampling}, the phenomenology of  extreme drag fluctuations is investigated based on direct sampling.
Both the instantaneous drag and time-averaged drag are considered.
It is found in particular that sampled extreme fluctuations of the instantaneous drag result from very similar dynamics. 
Section~\ref{sec:rare_events_algorithms} examines the applicability of both the \ac{ams} and \ac{gktl} algorithms to the simulation of extreme drag fluctuations in the same flow configuration.
In subsection~\ref{sec:ams}, we show that the use of the \ac{ams} algorithm is not successful, or at least not straightforwardly. This difficulty is put in perspective with the phenomenology of extreme drag fluctuations developed in  section \ref{sec:direct_sampling}.
Subsection~\ref{sec:gktl} presents the computation of extremes of the time-averaged drag by using the \ac{gktl} algorithm.
This latter allows for an exceptional reduction of the computational cost needed to simulate trajectories corresponding to extreme values of the time-averaged drag.
As a specific successful application, the \ac{gktl} algorithm is used to compute the return times of extreme fluctuations of the time-averaged drag acting on the immersed obstacle.
Perspectives and conclusion end this work.

\section{Description of the numerical case study}
\label{sec:test_flow}

\begin{figure}
	\centering
	\includegraphics[width=\linewidth]{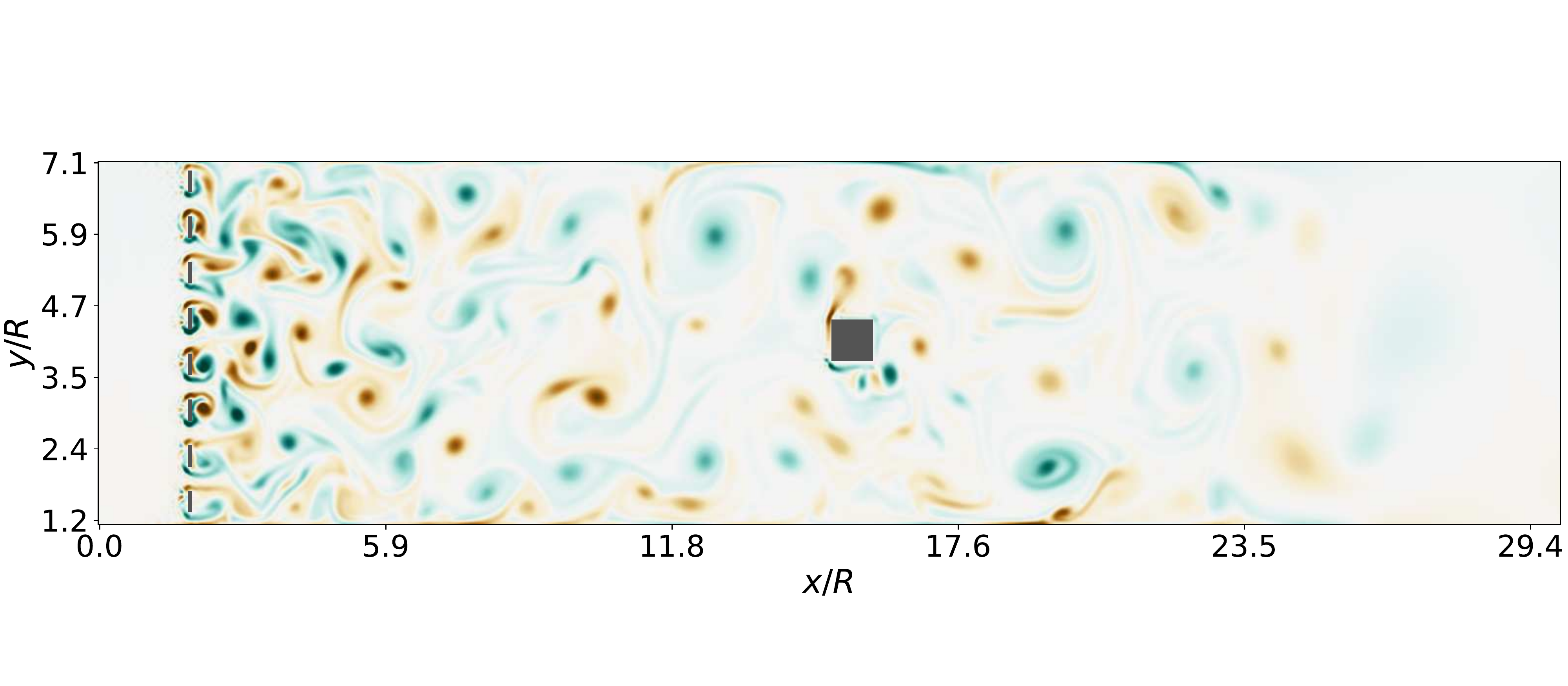}
	\caption{Our case study is a grid-generated turbulent flow impinging onto a fixed squared obstacle (of size $R$) located at the centre of a channel in two dimensions. The flow is artificially damped near the end of the channel. In the developed flow, turbulent eddies have typically the size of the square, which results in strong load fluctuations. The vorticity is displayed with an arbitrary colour map from blue (negative values) to red (positive values).}
	\label{fig:illustr_ecoulement}
\end{figure}

% introduce the flow 
%
The drag exerted by a grid-generated turbulent flow onto a fixed squared obstacle is considered as a representative case study (see Fig.~\ref{fig:illustr_ecoulement}). 
% why this flow?
%
Although real-world applications would eventually imply three-dimensional dynamics, a simplified two-dimensional setting has been chosen here to reduce the computational cost and allow for a systematic study. 
We believe that this system embeds the characteristic features that make
this study both relevant and challenging for fluid-structure-interaction problems. 

% rapid description of the flow
Turbulent eddies generated in the near-wake of the grid are carried downstream.
They interact with each other and grow in size as expected for two-dimensional turbulent dynamics. 
The dimension of the grid is such that the size of the eddies that hit the square is comparable to its size, resulting in strong fluctuations of the drag acting on the square. 
% lattice Boltzmann method 
%
The flow dynamics is integrated numerically by the \ac{lbm} \citep{kruger_lattice_2017}. 
While traditional methods in computational fluid dynamics rely on a discretization of the Navier-Stokes equations, the \ac{lbm} considers the fluid at an underlying  mesoscopic level.
Capturing the dynamics of collections of fluid particles moving and colliding on a lattice is here preferred to solving nonlinear \acp{pde}. 
%This seems crazy, however, most details at the mesoscopic level play actually no role at the macroscopic level. Therefore, the LB algorithm may be viewed as a minimal kinetic scheme compliant to the fluid dynamics at the macroscopic level.
Further details about the \ac{lbm} are given in Appendix \ref{app:lbm} and references therein.
In our context, this numerical method has been chosen principally for its computational efficiency.

% geometry
%
The simulated flow develops in a long plane channel of dimension $513 \times 129$ mesh points. The square obstacle has size $R=16$ (in mesh units) and is located at the centre of the channel. The spacing and bar height of the entrance grid are both equal to $R/2$ (see Fig.~\ref{fig:illustr_ecoulement}). 
% boundary conditions
%
No-slip boundary conditions are enforced on top and bottom walls of the channel and on the surface of the obstacle by using an halfway bounce-back procedure~\citep{lbm_book}.  
Upstream of the grid, a constant parabolic velocity profile and a constant mass density (equal to unity) are imposed as an inlet condition. 
The centreline velocity is $0.05$ in lattice units, \textit{i.e.} normalised by $\Delta x$ and $\Delta t$ referring to the lattice resolution and the time-step respectively. The initial distributions are imposed at equilibrium (see Appendix \ref{app:lbm}). 
In the bulk, the viscosity is adjusted so that grid turbulence is generated with Reynolds number $\mathrm{Re_{grid}}=1200$. The reference Mach number is equal to $0.06$ in agreement with the assumption of weak compressibility of the \ac{lbm}. 
Near the end of the channel, the flow is progressively damped within a \textit{sponge zone} where the viscosity is artificially enhanced.
Finally, the outlet boundary condition relies on a second-order extrapolation of the velocity and mass density.
The extrapolated distributions are evaluated through a regularization procedure relying on a finite difference estimation of the local stress tensor, as introduced in \citep{latt2008straight}. For more details about the implementation of the flow, see the software repository attached to this paper (\url{https://github.com/tlestang/paper_extreme_drag_fluctuations}).

\subsection{The drag force}
\label{sec:drag_force}

\begin{figure}
	\centering
	\includegraphics[width=\linewidth]{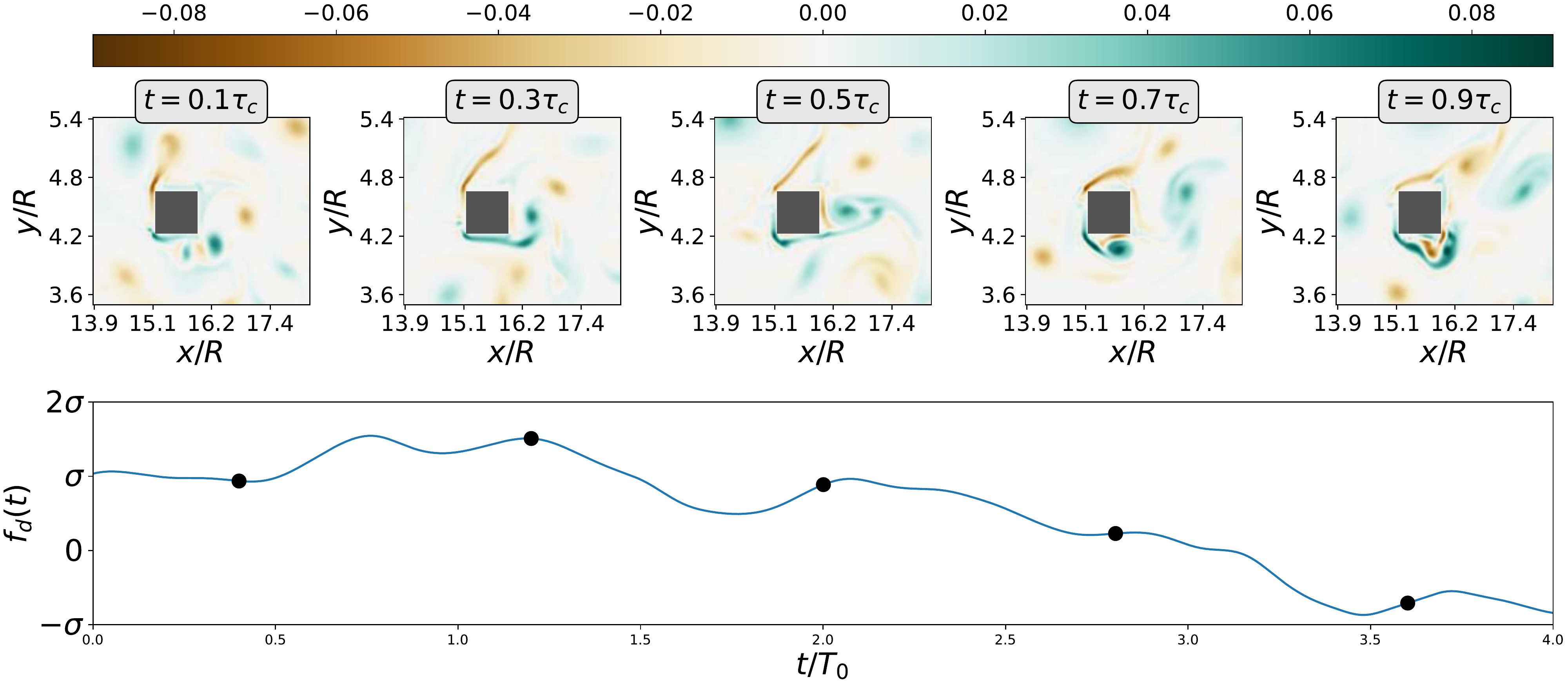}
	\caption{Snapshots of the vorticity related to typical drag fluctuations (within one standard deviation) over a time interval of length $\tau_c \simeq 4T_0$; $\tau_c$ will later be identified as the correlation time of the drag signal.
		The vorticity is given in lattice units.}
	\label{fig:typical_vorticity}
\end{figure}

%% LB parameters  %

% drag signal
%
The incoming turbulent flow exerts a fluctuating force onto the squared obstacle.
The \textit{drag} is defined as the resulting force in the streamwise $x$-direction. Formally 
\begin{equation}
\label{eq:drag_definition}
f_d(t) = \int_{\mathcal{S}} \boldsymbol{\tau}_{x \beta}(\mathbf{x},t) ~ \mathrm{d}{\mathcal{S}}_\beta(\mathbf{x}),
\end{equation}
where $\mathcal{S}$ is the surface of the obstacle and $\boldsymbol{\tau}$ denotes the stress tensor (see Appendix \ref{app:lbm}). 
Here, the viscous stress makes a negligible contribution to the drag. The latter therefore results mostly from pressure forces.
Since the pressure on the top and bottom sides of the square applies in the normal direction, they do not contribute to the drag. 
As a consequence, the drag can eventually be expressed as the difference 
\begin{equation}
\label{eq:drag_approx}
f_d(t) = p_{fb}(t) - p_{base}(t)
\end{equation}
between the pressure integrated over the upstream side of the obstacle or \textit{forebody}, $p_{fb}(t)$, and the downstream side or \textit{base} $p_{base}(t)$.
Pressure fluctuations are related to the dynamics of the vorticity field.
Regions of strong vorticity correspond to strong local pressure gradients, \emph{e.g.} as demonstrated analytically with a Rankine vortex.
%

% Definition of turnover time
%
The typical timescale (turnover time) of drag fluctuations can be estimated from dimensional analysis as
\begin{equation}
\label{eq:turnover_time}
T_0 = \frac{R}{U},
\end{equation}
where $R$ is the size of the square and $U$ is the averaged velocity in the channel. 
Fig.~\ref{fig:typical_vorticity} displays the evolution of the vorticity field around the obstacle for typical drag fluctuations.
%\sout{Because the vorticity generated along the forebody is swept away by the mean flow, the pressure field in the vicinity of the base is only slightly perturbed.}
Vorticity is generated along the forebody and eventually carried away by the flow. Typical fluctuations of the drag (within one standard deviation) do not result from some preferred arrangement of the vorticity around the obstacle.

\subsection{The drag as a random process
	%: Probability Density Function and correlation in time
}
\label{sec:pdfs}

\begin{figure}
	\centering
	\includegraphics[width=\linewidth]{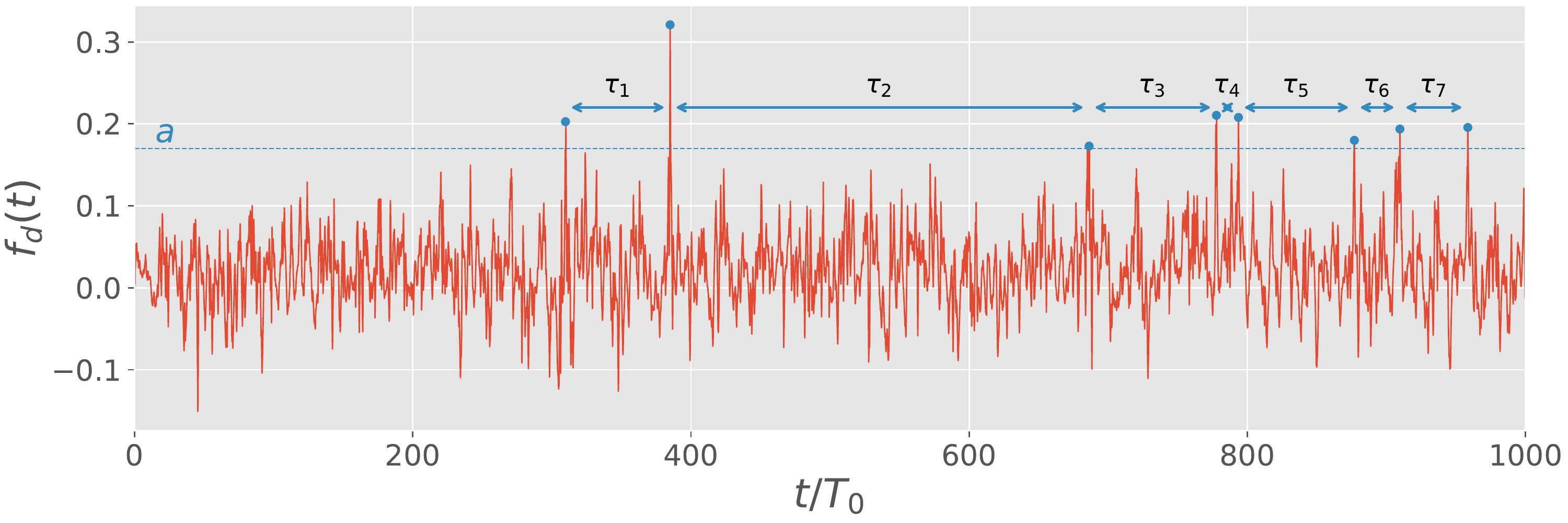}
	\caption{
	Temporal evolution of the drag (in lattice units) acting on the square under the action of the impinging turbulent flow. The time has been normalised by the turnover time related to the mean-flow velocity and the size of the obstacle, \emph{i.e.} $T_0=R/U$. The \textit{return time} $r(a)$  is the averaged waiting time between the occurrence of peak fluctuations of amplitude larger than $a$. If the threshold $a$ is sufficiently high, one observes that $r(a)$ is much larger than the correlation time of the signal. The selected peak fluctuations are therefore well separated.}
		\label{fig:illustrate_return_time}
\end{figure}

% drag signal
%
% Description du signal de trainee typique
In Fig.~\ref{fig:illustrate_return_time},
the time signal of the drag acting on the square, $f_d(t)$, appears unpredictable in details and exhibits repeated bursts of high amplitude that deviate significantly from the averaged value.
Therefore, it is natural to model the drag as a (scalar) random process.

\begin{figure}
	\centering
	\subfloat[\ac{pdf} of normalised drag fluctuations]
	{\label{fig:pdf_drag_a}
		\includegraphics[width=.4\linewidth]{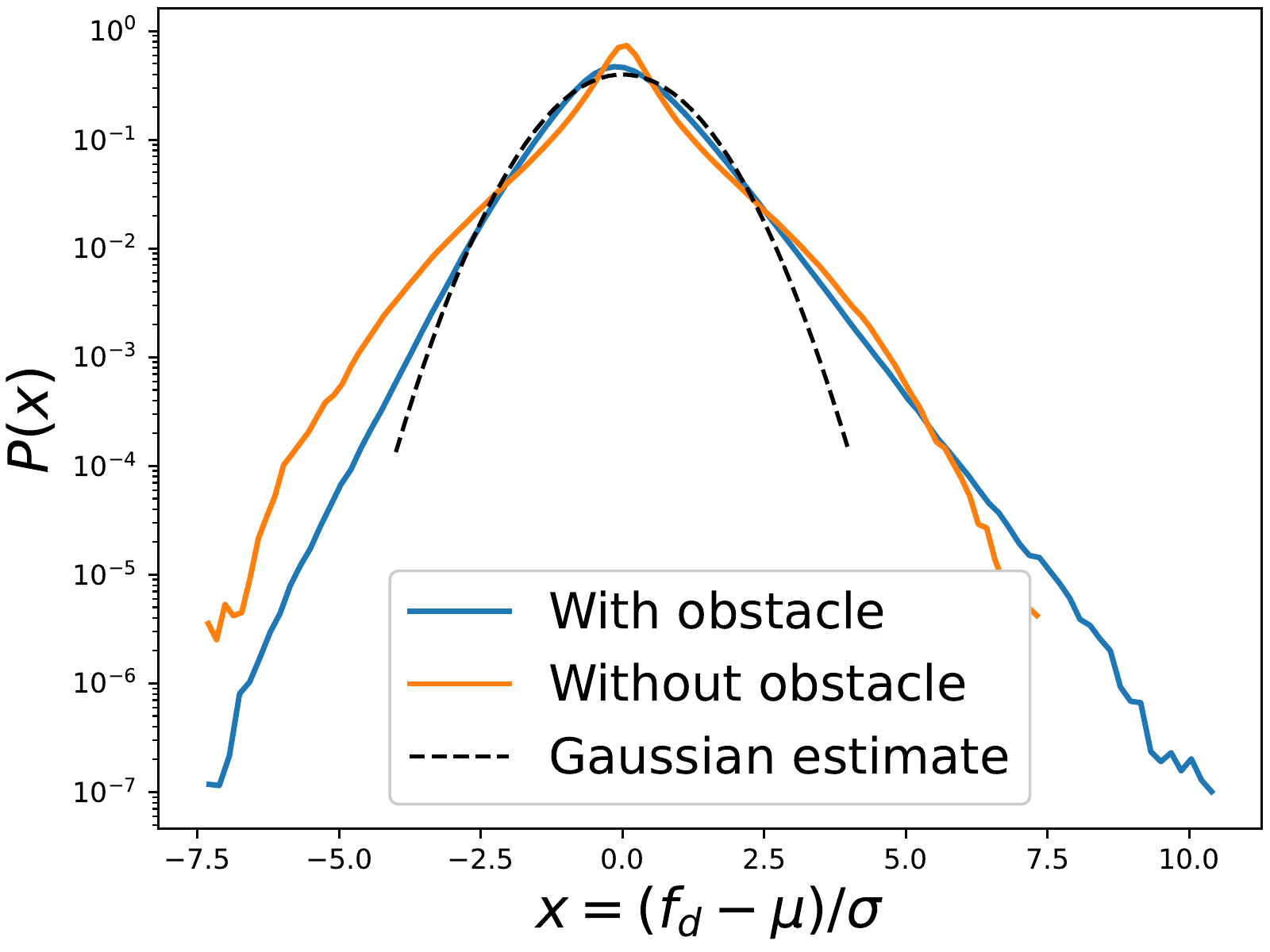}}
	\subfloat[Autocorrelation of drag fluctuations]
	{\label{fig:pdf_drag_b}
		\includegraphics[width=.45\linewidth]{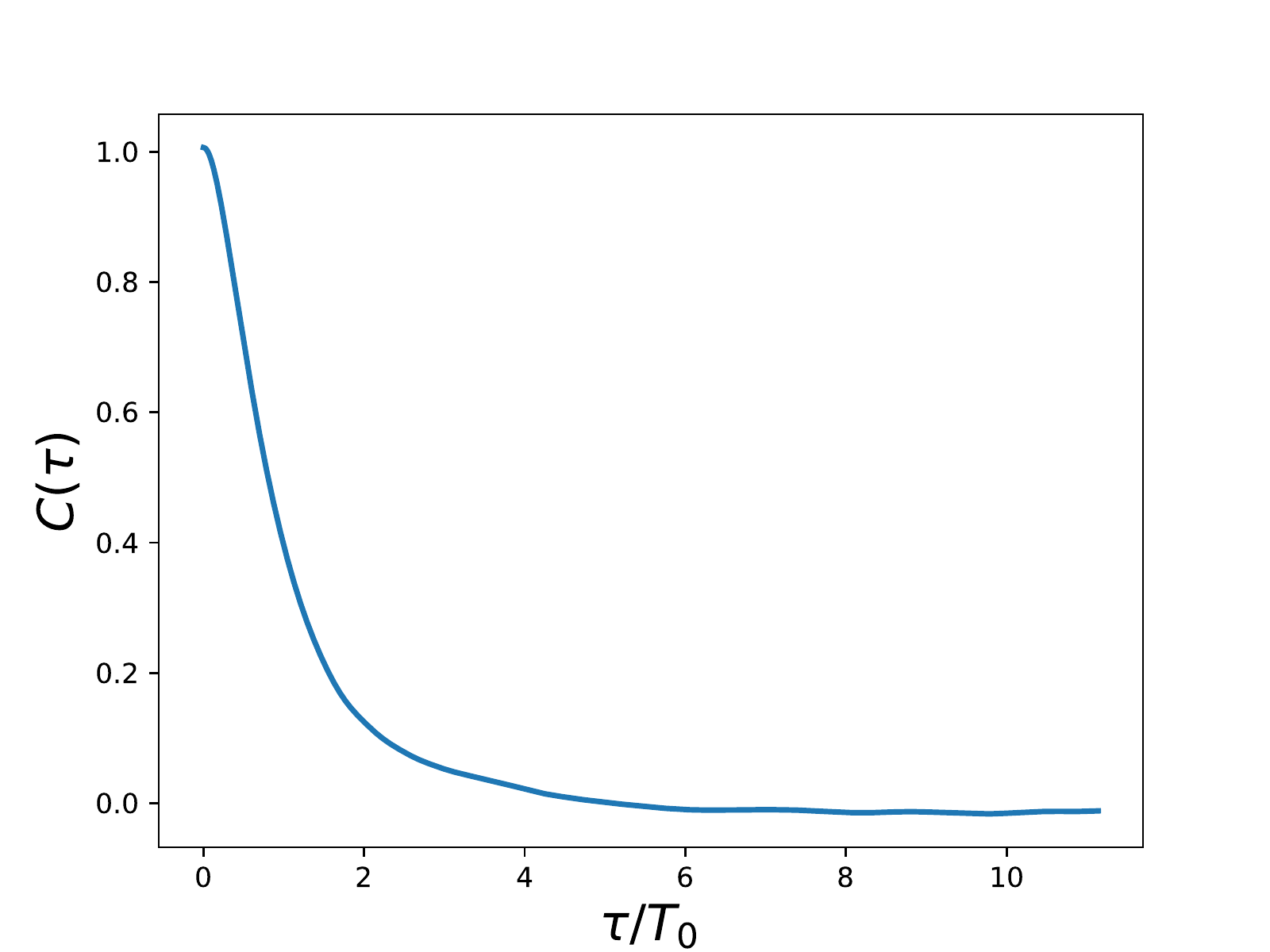}}  
	\caption{{(a)} \ac{pdf} of normalised drag fluctuations ($\tilde f_d = f_d - \bar{f_d}$) where $\mu = \bar{f_d}$ denotes the time-averaged value and $\sigma = \sqrt{\overline{{\tilde f_d}^2}}$ is the standard deviation. The drag is evaluated both in the presence (blue) and in the absence (red) of the obstacle. %Note that the amplitudes have not been normalized.
		{(b)} Autocorrelation function of the drag defined as $C(\tau) = \overline{ \tilde f_d(t+\tau)\tilde f_d(t)} ~/~ \overline{{\tilde f_d}^2}$. The correlation time $\tau_c\simeq 4 T_0$ is given by $C(\tau_c)=0$.
		%$\tau_c$ may therefore be considered as the correlation time of the drag signal.
	}
	\label{fig:pdf_drag}
\end{figure}

% drag statistics
% pdf
Drag fluctuations have been sampled along a simulation of duration $T_{tot} = 4\times 10^6~T_0$.
This long simulation will be referred to as the \textit{control run} in the following.
It has been made possible by the relative simplicity of the investigated flow and the computational efficiency of the \ac{lbm}.
The \ac{pdf} of drag fluctuations is shown in Fig.~\ref{fig:pdf_drag_a}.
It deviates from a normal law and shows an exponential tail for large positive fluctuations, \textit{i.e.} ${\mathbb{P}}(f_d) \propto e^{-\ell f_d}$.
Fig.~\ref{fig:pdf_drag_a} also displays the \ac{pdf} of drag fluctuations acting on a control surface corresponding to the periphery of the obstacle but in the absence of the obstacle. 
In that case, the \ac{pdf} is quasi-symmetric and does not display exponential tails. This shows that the asymmetry of the \ac{pdf} and the development of a positive exponential tail are closely related to the no-slip condition on the obstacle boundary.

% drag statistics
% correlation time
Lastly, the autocorrelation function of the drag $C(\tau)$ is shown in Fig.~\ref{fig:pdf_drag_b}. It is found that drag fluctuations are correlated over a time interval $\tau_c \simeq 4T_0$, illustrating that the drag loses its memory  over a time scale corresponding to the sweeping of a few eddies past the obstacle.
This observation is important for the application of rare-event algorithms as it will be discussed in section~\ref{sec:rare_events_algorithms}.
In the following, $\tau_c$ will be referred to as the \textit{correlation time} of the drag process.
The ratio $T_0 / \tau_c$ may be viewed as a {Strouhal number}. The value $St=0.25$ is consistent with common observations for flows past blunt structures at comparable Reynolds numbers \citep{rodi1998}.
\section{Extreme fluctuations of the drag by means of direct sampling}
\label{sec:direct_sampling}

\begin{figure}
	\centering
	\includegraphics[width=.6\linewidth]{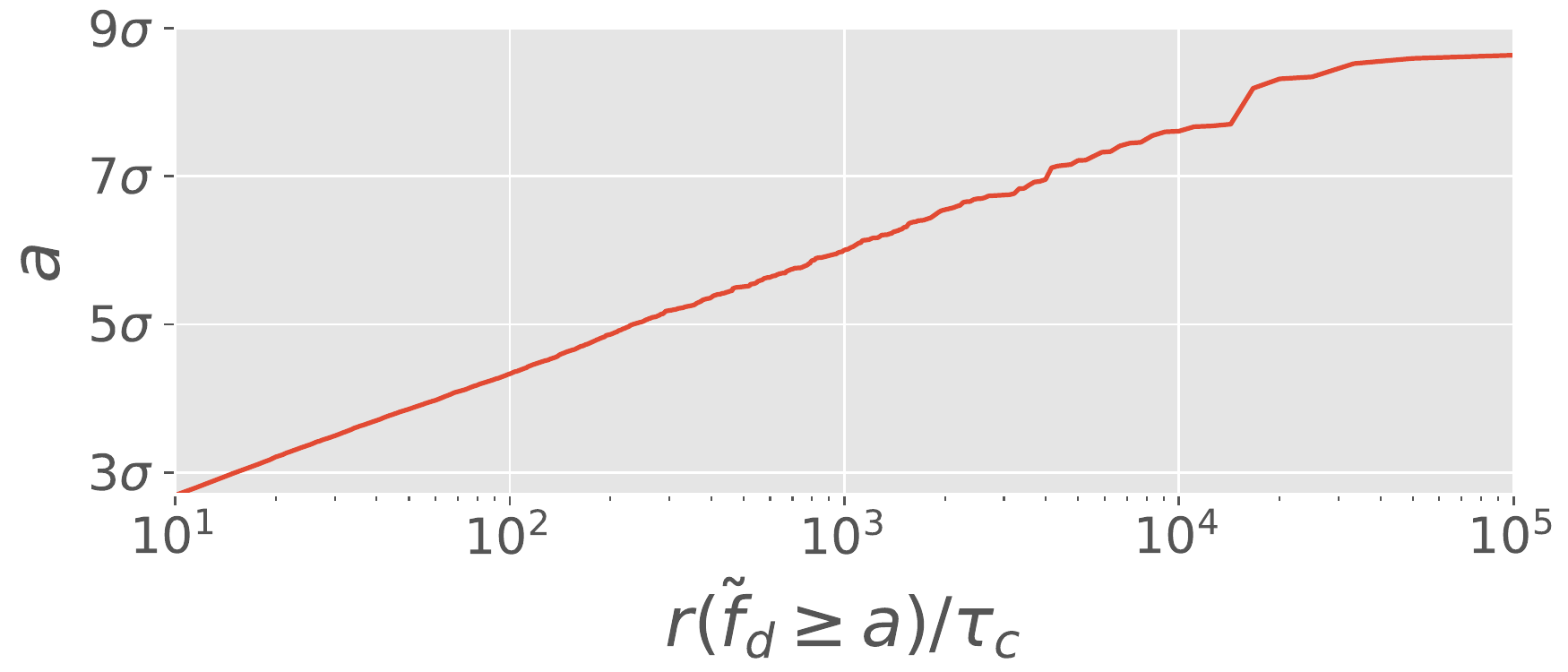}
	\caption{Amplitude of drag fluctuations as a function of the corresponding return time. $\tilde{f}_d$ denotes the drag with zero mean, \textit{i.e.} $\tilde{f}_d = f_d - \overline{f_d}$.
	}
	\label{fig:return_time_instant}
\end{figure}

% return time given from direct sampling
%
The phenomenology of extreme fluctuations of the drag is first investigated through brute-force direct sampling applied to the control run.
Direct sampling is here used as opposed to approaches involving rare-events algorithms discussed in section~\ref{sec:rare_events_algorithms}.
It will provide a trustworthy baseline for the validation of rare-events algorithms. 

The waiting times $\tau$ are defined as the time between two consecutive occurrences of peak fluctuations with amplitude $f_d \geq a$, as illustrated in Fig.~\ref{fig:illustrate_return_time}.
The mixing time $\tau_m$ is the time needed for the dynamics to lose the memory of its initial condition.
As soon as the typical waiting times are much larger than the mixing time $\tau_m$, the occurrences of such events follow a Poisson process and the distribution of the waiting times is exponential, \emph{i.e.} $P(\tau)=\lambda(a)\exp(-\lambda(a)\tau)$ where $r(a)=1/\lambda(a)$ is the averaged waiting time \citep{lestang_computing_2018}; $r(a)$ is called the {\it return time} of the level $a$.
For systems without multi-stability, it is common for the mixing time $\tau_m$ to be of the order of the correlation time $\tau_c$.

How rare is a fluctuation $a$ is quantified by the return time $r(a)$.
We can define extreme drag fluctuations as \textit{rare events} in the sense that the return time is much larger than the correlation time, \emph{i.e.} $r(a) \gg \tau_c$.

If one assumes that  $r(a) = t(a)~/~\mathbb{P}(f_d\geq a)$   where the time scale $t(a)$ is of order $\tau_c$ and varies much more slowly with $a$ than ${\mathbb{P}(f_d\geq a)}$,
one might expect that 
\begin{equation}
  \label{eq:return_time}
  r(a) \underset{a\to\infty}{\propto} \exp(\ell a)
\end{equation}
where $l$ is the rate describing the positive tail of the \ac{pdf} of the drag (shown in Fig.~\ref{fig:pdf_drag}).
Fig.~\ref{fig:return_time_instant} shows the evolution of the return time $r(a)$ with the amplitude of fluctuation $a$, computed from {direct sampling} of the drag signal $f_d(t)$~\citep{lestang_computing_2018}.
Consistently, it is found that the return time $r(a)$ is well approximated by an exponential for large levels $a$. Let us also point out some deviation from the exponential law at the largest levels, which is probably the consequence of under-sampling.

\subsection{Extracting extreme drag fluctuations from a very long timeseries}
\label{sec:extreme_extraction}

% intro
%
We have extracted the fluctuations of the drag with a return time $r(a)$ greater than  $10^4\tau_c$ from the control time-series $\{f_d(t)\}_{0 \leq t \leq T_{tot}}$.
This set will be considered as representative of \emph{extreme events} in the upcoming study.
The choice of this particular threshold has been driven by the need to collect enough events with large amplitude and to possibly identify generic features.
According to Fig.~\ref{fig:return_time_instant}, the related amplitude $a$ is found equal to $7.6~\sigma$ with $\sigma =\sqrt{\overline{\tilde{f_d}^2}}$ being the standard deviation of the drag process, where $\tilde{f_d}$ denotes the drag with zero-mean, \textit{i.e.} $\tilde{f}_d = f_d - \overline{f_d}$.
Precisely, 104 independent fluctuations with $\tilde{f}_d(t) \geq 7.6\sigma$ have been identified. Each fluctuation is characterized by its maximal value, $f_d^{\star}$, and the time, $t^{\star}$, at which this maximum is reached.
In the following, the phenomenology of extreme drag fluctuations will be examined on the basis of this set of events.

\subsection{Extreme fluctuations of the instantaneous drag}
\label{sec:instantaneous_drag}

\subsubsection{Contribution of forebody and base pressure fluctuations to the overall drag fluctuation}
\label{sec:forebody_and_base_contribution}

\begin{figure}
	\centering
	\includegraphics[width=.8\linewidth]{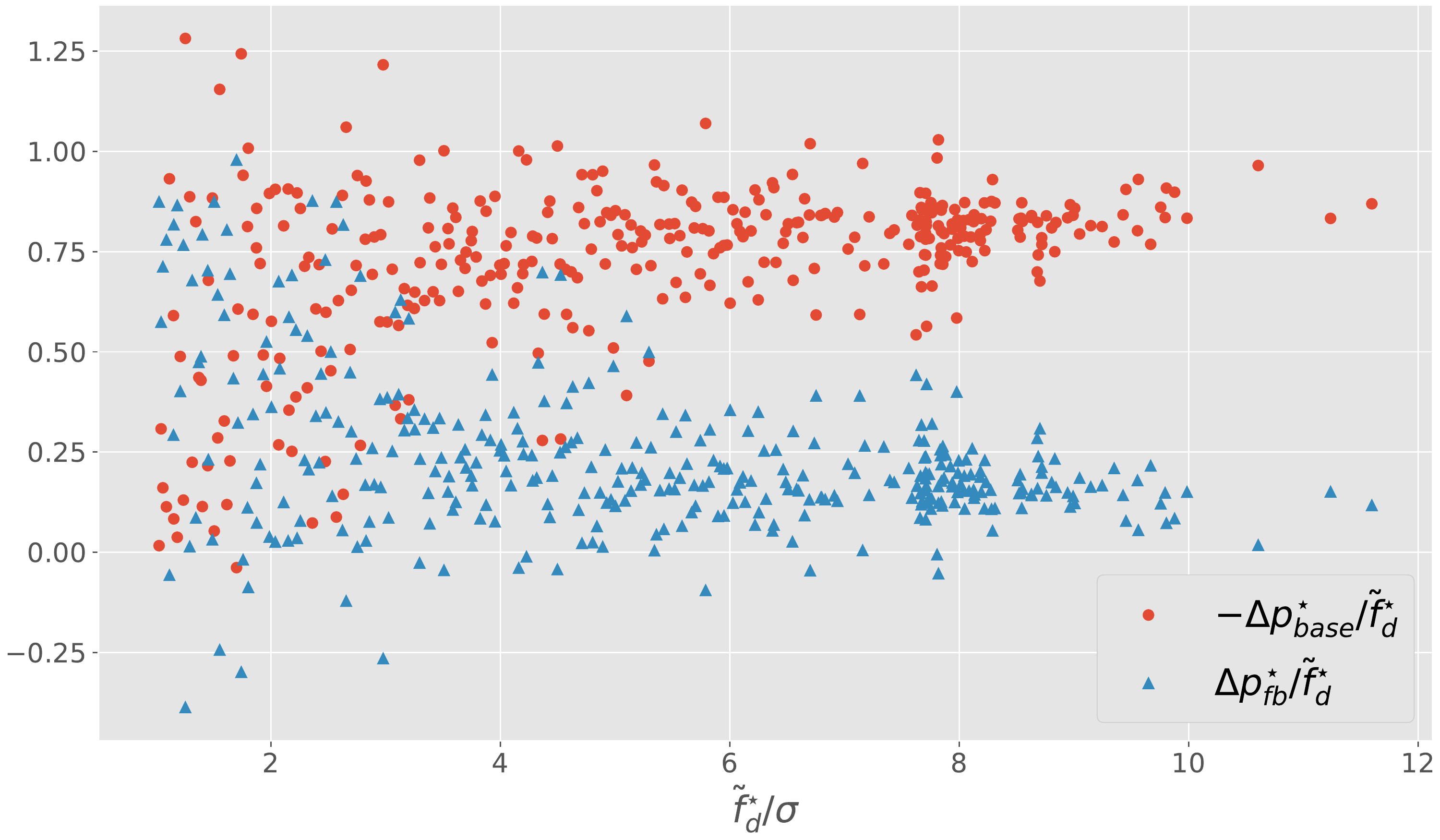}
	\caption{\label{fig:pressure_ratio} Relative contributions of the forebody and base pressure variations to extreme amplitudes of the drag fluctuation $\tilde f_d$. An extreme event corresponds to an amplitude $\tilde f^{\star}_d$ and a unique pair  ($\tilde{p}^{\star}_{base}$,~$\tilde{p}^{\star}_{fb}$).}
\end{figure}
In section~\ref{sec:test_flow}, it was pointed out that drag fluctuations within one standard deviation were not associated with any particular arrangement of the   vorticity around the obstacle. We shall see in the following that the situation is different in the case of {extreme} drag fluctuations.
%
% relative contribution of forebody and base pressure
%
Let $(t^{\star}, f_d^{\star})$ refer to an extreme-drag event.
The (zero-mean) fluctuation $\tilde{f}_d^{\star} = f_d^{\star} - \overline{f_d}$ can be  decomposed into 
\begin{equation}
\tilde{f}_d^{\star} = \Delta p_{fb}^{\star} - \Delta p_{base}^{\star}
\end{equation}
where $\Delta p_{fb}^{\star}$ and $\Delta p_{base}^{\star}$ denote the variations of the forebody and base pressure, respectively.
Fig.~\ref{fig:pressure_ratio} displays the relative contributions
$\Delta p_{fb}^{\star}/\tilde{f}_d^{\star}$ and $-\Delta p_{base}^{\star}/\tilde{f}_d^{\star}$ to the drag fluctuation $\tilde f_d^{\star}$.
	It is found that very large fluctuations of the drag result mainly ($\sim 80\%$) from a drop in base pressure, whereas the variation of forebody pressure contributes much less to $\tilde{f}_d^{\star}$. On the contrary, moderate fluctuations arise from combined variations of the forebody and base pressures without any particular predominance, which is in agreement with the previous observations (see Fig.\ref{fig:typical_vorticity}).

\subsubsection{Fluid dynamics related to extreme drag fluctuations}
\label{sec:dynamical_aspects}

\begin{figure}
	\centering
	\includegraphics[width=.7\linewidth]{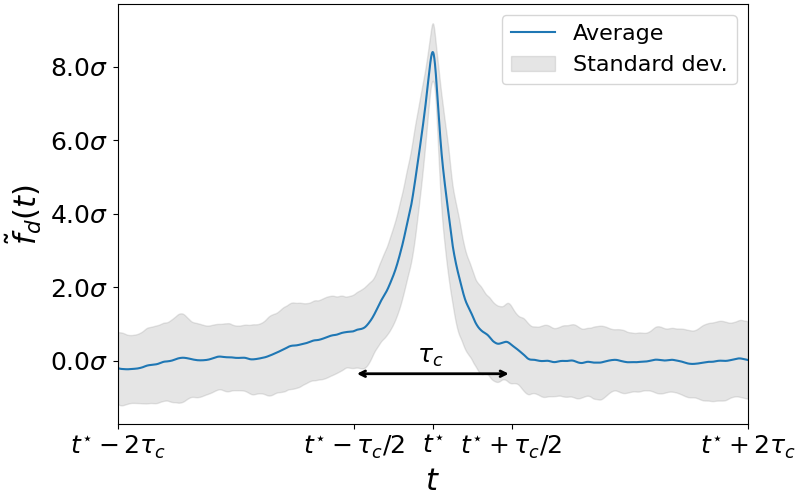}
	\caption{\label{fig:timeseries_extremes} Ensemble average of drag signals centred around extreme fluctuations occurring  at $t=t^{\star}$. The blue line shows the mean profile whereas the shaded area indicates variations (around the mean profile) within one standard deviation. Extreme-drag events exhibit a typical lifetime of one correlation time $\tau_c$. The profile is slightly skewed indicating that the step up is slower than the return to typical values.}
\end{figure}

% mean profile around burst
%
The focus is now on the flow scenarii that yield extreme values of the drag. 
Fig.~\ref{fig:timeseries_extremes} displays the mean profile (in time) of the drag signal around extreme events. A peaked profile is observed with a width roughly corresponding to one correlation time $\tau_c$. This shows that the duration of extreme events corresponds typically to the sweeping time of the flow past the obstacle. 
Interestingly, the profile is also slightly skewed, indicating that the step up of the drag is slower than the return to typical values past the peak value.
This asymmetry (under time reversal) is closely linked to the symmetry breaking in what happens before and after the obstacle.
To better understand the flow scenarii leading to these events, the vorticity fields around the obstacle are now examined.

\begin{figure}
	\centering
	\includegraphics[width=\linewidth]{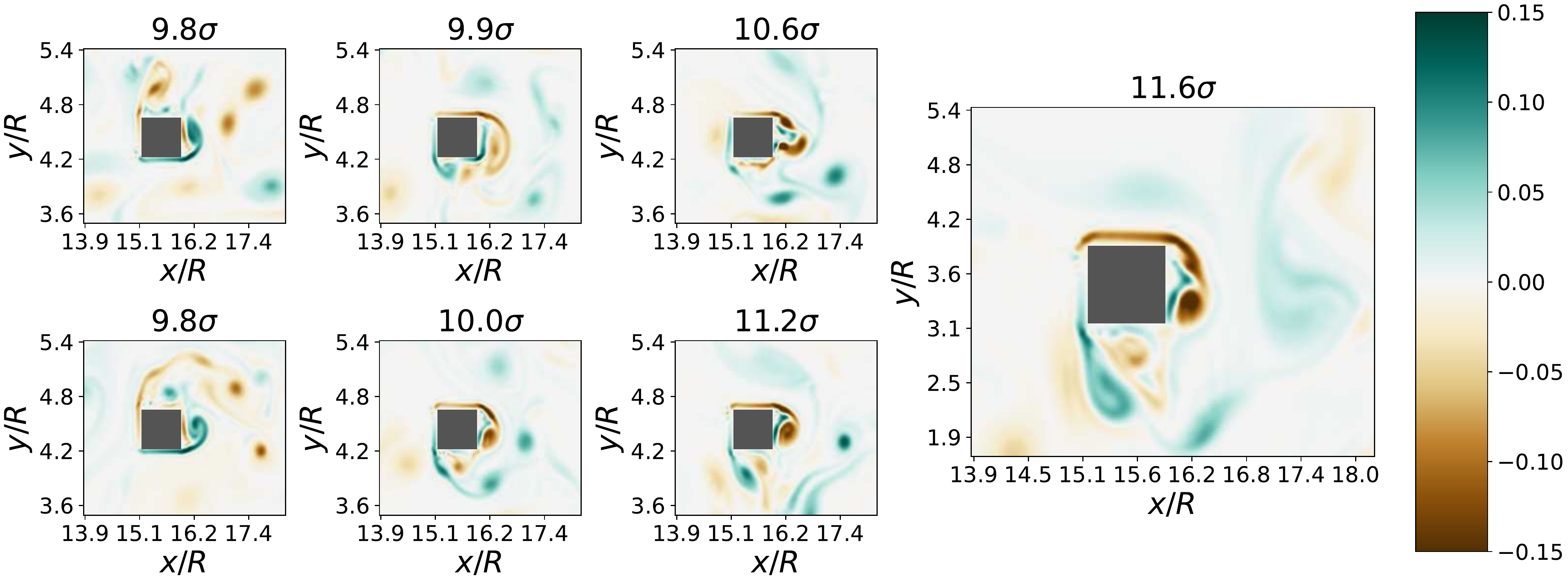}
	\caption{\label{fig:top_4_events_vorticity} Vorticity field (in lattice units) around the obstacle at $t=t^{\star}$ for the highest drag amplitudes recorded in the control run.
	}
\end{figure}

\begin{figure}
	\centering
	\includegraphics[width=.5\linewidth]{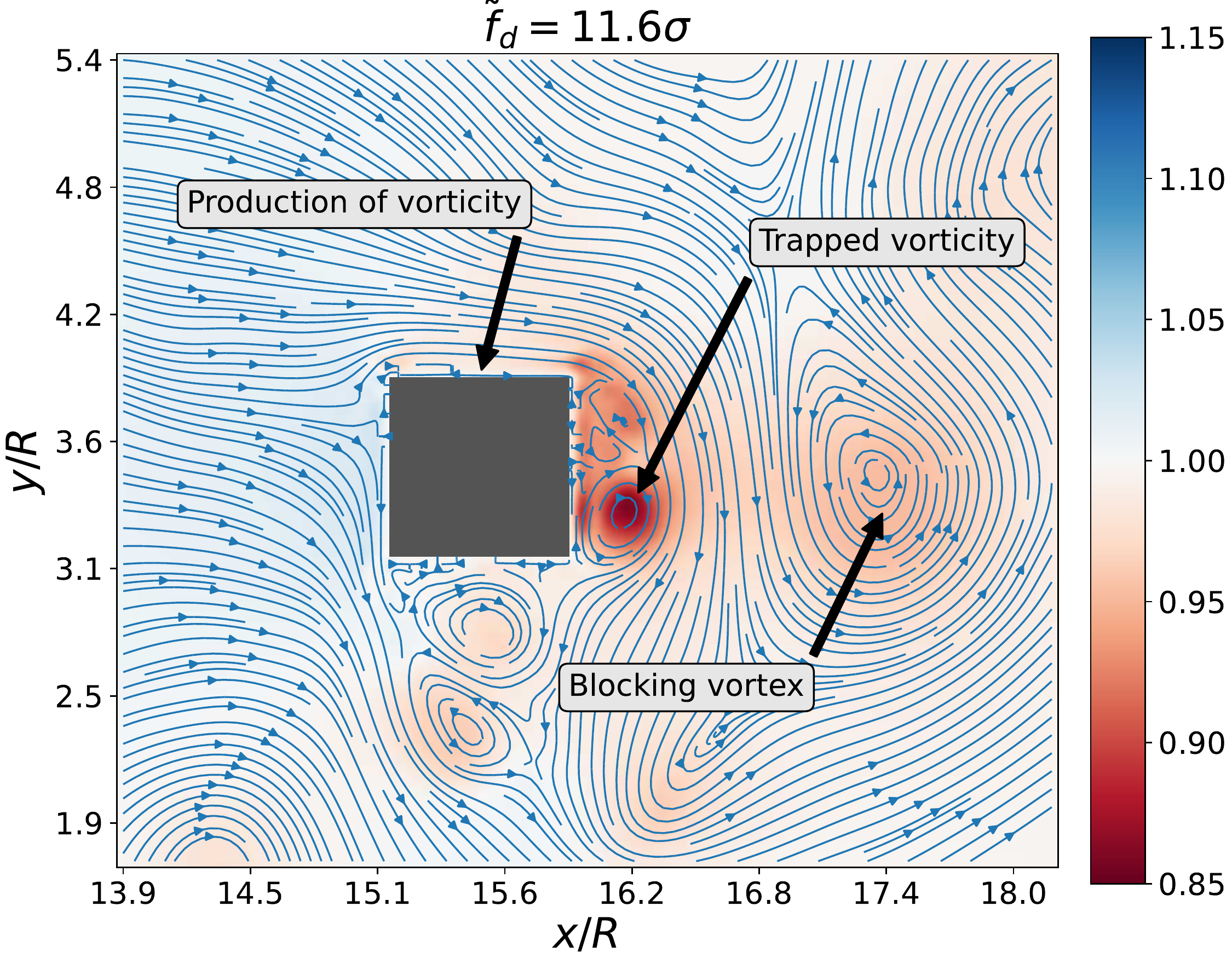}
	\caption{\label{fig:density+streamlines} Pressure field (in lattice units) and velocity streamlines at $t=t^{\star}$. In the near wake of the square, a (blocking) vortex blocks an intense vortex against the base of the obstacle.}
\end{figure}

\begin{figure}
	\centering
	\includegraphics[width=.8\linewidth]{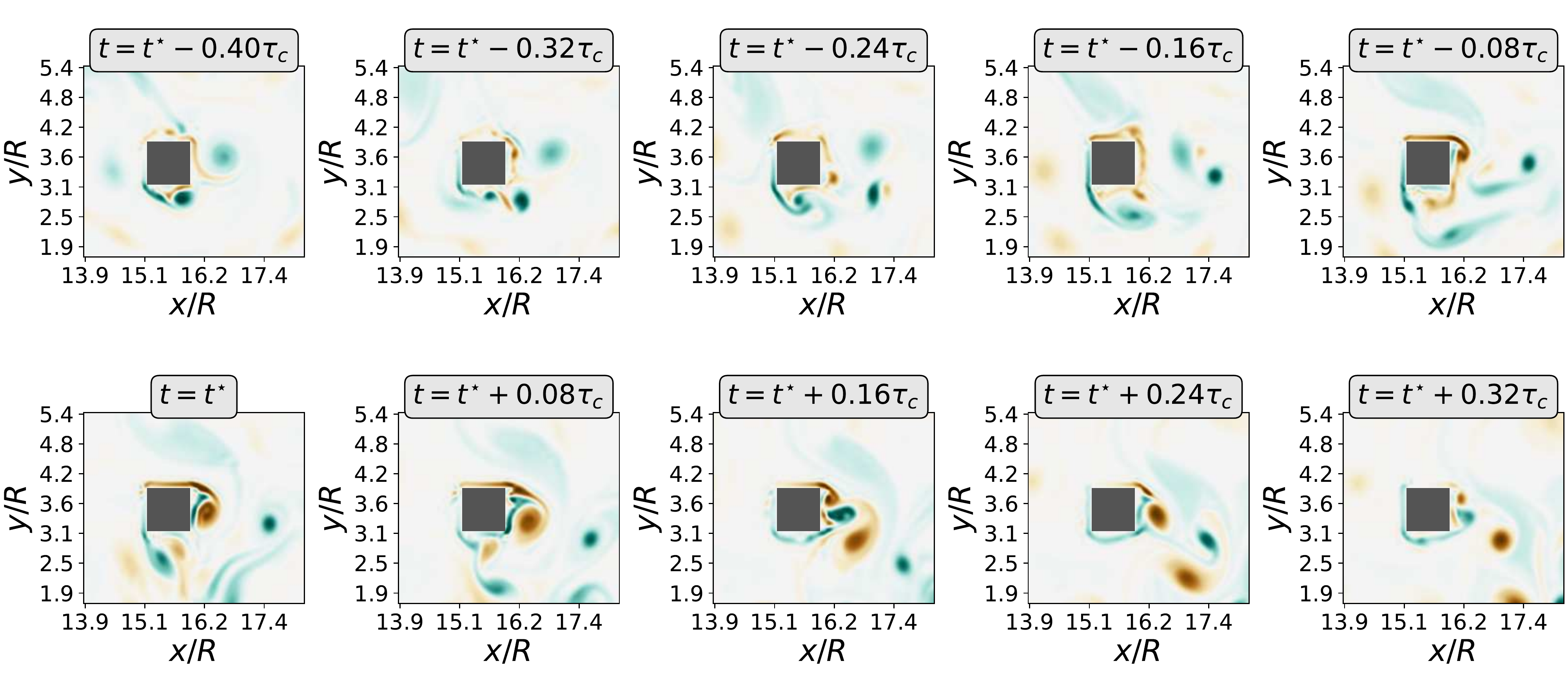}
	\caption{\label{fig:vorticity_dynamics} Snapshots of the vorticity field (in lattice units)  around $t=t^{\star}$.}
\end{figure}
% flow scenario
%
Fig.~\ref{fig:top_4_events_vorticity} displays the vorticity field (in lattice units) around the obstacle for the highest amplitudes of the drag during the control run.
In each case, an intense vortical structure is visible near the base of the obstacle.
The vorticity level of this structure is typically twice the amplitude of typical vorticity fluctuations observed in Fig.~\ref{fig:typical_vorticity}.
%+
The formation of this vortex originates from an intense negative (or positive) vorticity layer at the top (or bottom) boundary of the obstacle. 
This high vorticity is responsible for a significant pressure drop at the base of the obstacle and therefore a strong drag.
In contrast, nothing special happens near the forebody of the obstacle during extreme-drag events.  

The high pressure drop near the base of the obstacle appears to be closely related to the presence of a strong vortex blocked against the base. 
As illustrated in Fig.~\ref{fig:density+streamlines}, this blockage is enforced by the presence of opposite vorticity in the near wake, which holds the vortex against the base of the obstacle and prevents it from being swept away for a while.
This scenario is better evidenced by Fig.~\ref{fig:vorticity_dynamics}, where the time history of the vorticity field around $t=t^\star$ for the same event is shown. 
Before the occurrence of the extreme event, positive vorticity originating from the bottom boundary layer develops in the near wake of the square. This positive vorticity  prevents the shedding of negative vorticity and enforces the development of an intense vortex against the base of the square. 
As the blocking vortex is in turn advected downstream, the vortex against the base is released.
Consistently, one can argue that the typical duration of this scenario is related to the sweeping time of the flow past the obstacle, and is therefore of the order of $\tau_c$.
This is in full agreement with the typical duration obtained from statistical consideration on the mean profile of large-drag fluctuations in Fig.~\ref{fig:timeseries_extremes}.
This scenario is generic and has been observed for most extreme events sampled in the control run.

\begin{figure}
	\centering
	\includegraphics[width=.7\linewidth]{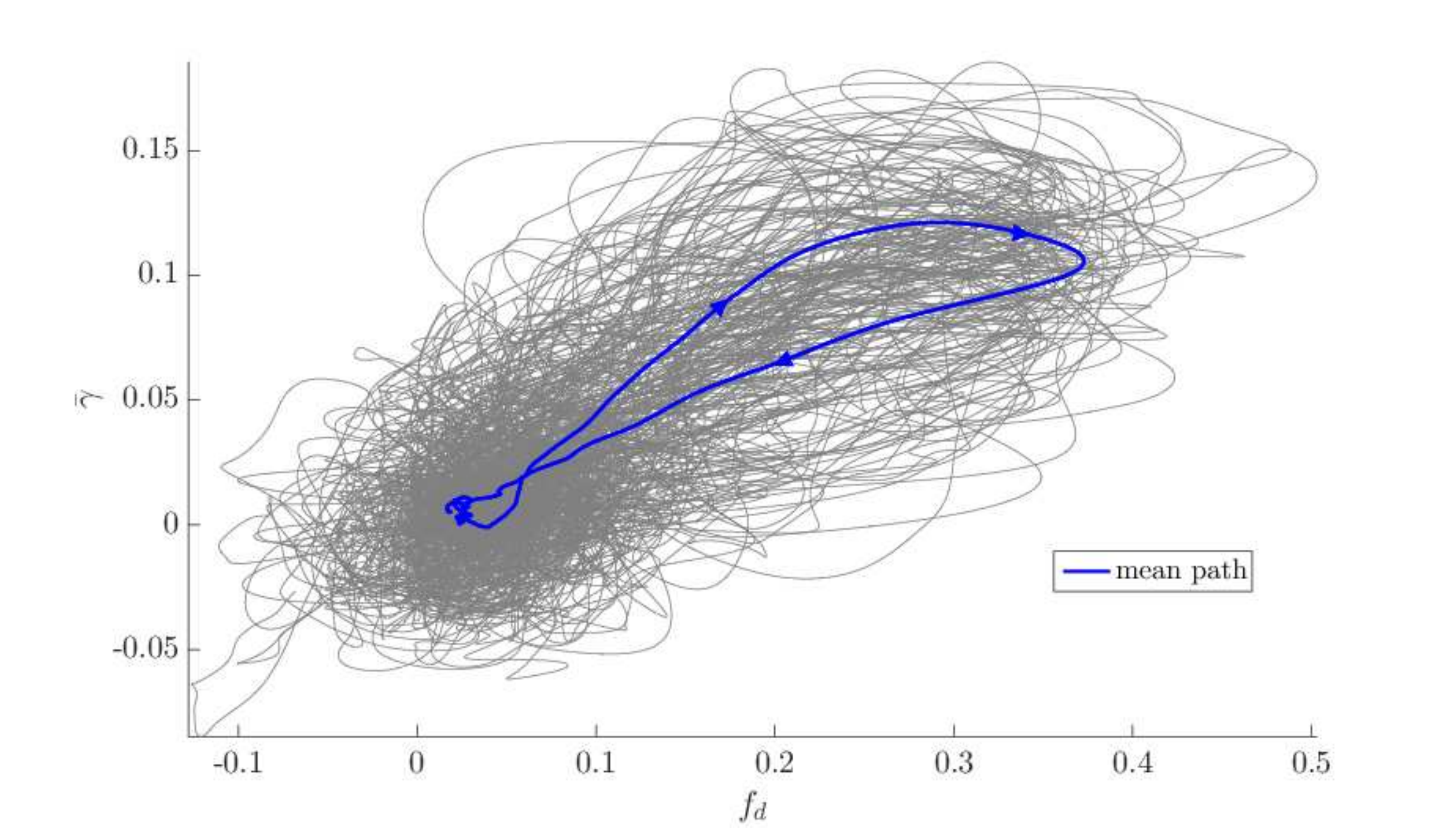}
	\caption{\label{fig:shear_asof_drag} Evolution of the (integrated) shear along the top or bottom sides of the obstacle as a function of the drag for $t^{\star}-2\tau_c \leq t \leq t^{\star}+2\tau_c$. Each trajectory corresponds to a single event. The blue line is the mean path averaged over the set of extreme events sampled in the control run.}
\end{figure}

Since the occurrence of large drag amplitudes arises from the production of vorticity along the top or bottom side of the square, it is proposed to characterize the dynamics of extreme events by their trajectory in the parameter space $(f_d(t), \bar{\gamma}(t))$ where $\bar{\gamma}(t)$ is the averaged shear along the top or bottom boundary of the square:
\begin{equation}
\label{eq:avg_shear_def}
\overline{\gamma} = \frac{1}{R} \int_{\mathcal{S}_\parallel} \frac{\partial u(\mathbf{x})}{\partial y}\mathrm{d}\mathbf{x},
\end{equation}
where $R$ denotes the size of the square, $u$ is the streamwise component of the velocity field and $\mathcal{S}_\parallel$ is the surface of either the top or the bottom boundary.
Fig.~\ref{fig:shear_asof_drag} shows $\overline{\gamma}(t)$ as a function of the instantaneous drag $f_d$(t) for $t^{\star}-2\tau_c \leq t \leq  t^{\star}+2\tau_c$ for the 104 sampled extreme events.
Before and after the extremal fluctuation, \textit{i.e.} for $t^{\star}-2\tau_c \leq t \leq t^{\star}-\tau_c$ and $t^{\star}+\tau_c \leq t \leq t^{\star}+2\tau_c$, paths wander in a region related to typical values of both $\overline{\gamma}$ and $f_d$.
On the contrary, the drag abruptly varies for $t^{\star}-\tau_c \leq t \leq t^{\star}+\tau_c$ near the extremal amplitude.
These excursions always go clockwise, that is, $\overline{\gamma}$ attains its maximum value before $f_d$ does. 
This is consistent with an increase of $\overline{\gamma}$ acting as a precursor for extreme drag fluctuations.
In this representation, we also observe that the path related to the increase of the drag is longer than the path related to the return to typical values, which may be related to the asymmetry of the mean profile displayed in Fig.~\ref{fig:timeseries_extremes}. 

\subsection{Extreme fluctuations of the time-averaged drag }
\label{sec:time_avg}

\begin{figure}
	\centering
	\includegraphics[width=\linewidth]{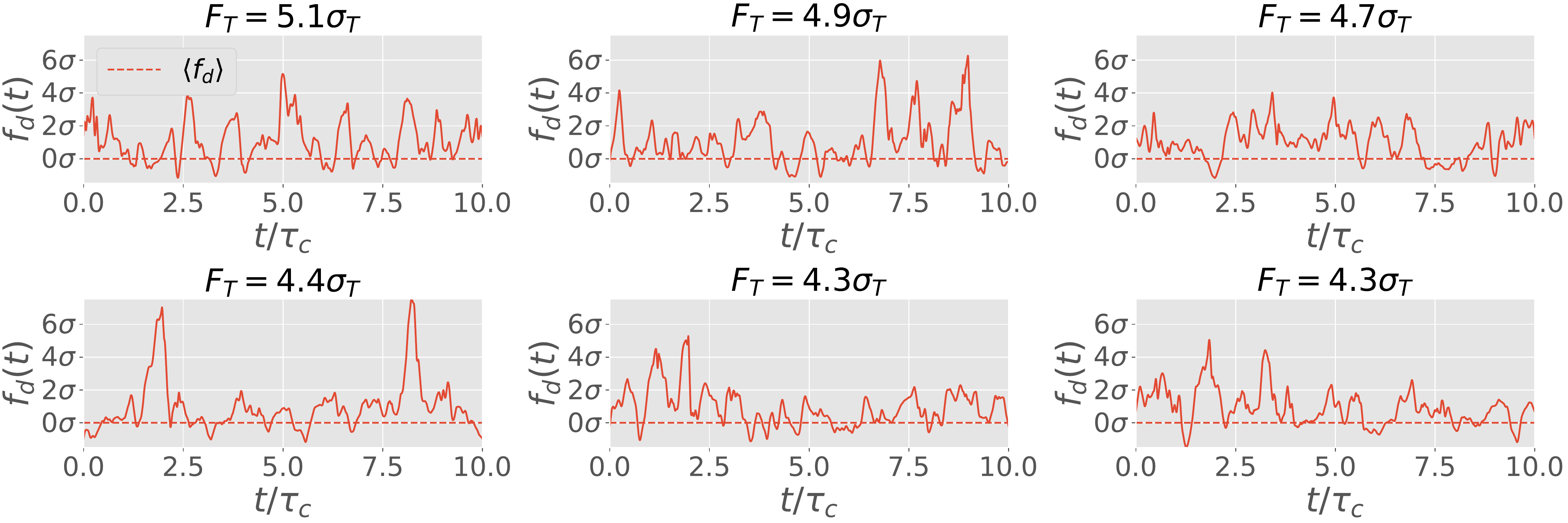}
	\caption{Instantaneous drag signals $f_d(t)$ corresponding to the highest fluctuations of the averaged  drag $F_T$ with a time window $T = 10 \tau_c$;  $\sigma$ and $\sigma_T$ denote the standard deviations of the instantaneous and the averaged drag, respectively.}
	\label{fig:extreme_avg}
\end{figure}

\begin{figure}
	\centering
	\includegraphics[width=.7\linewidth]{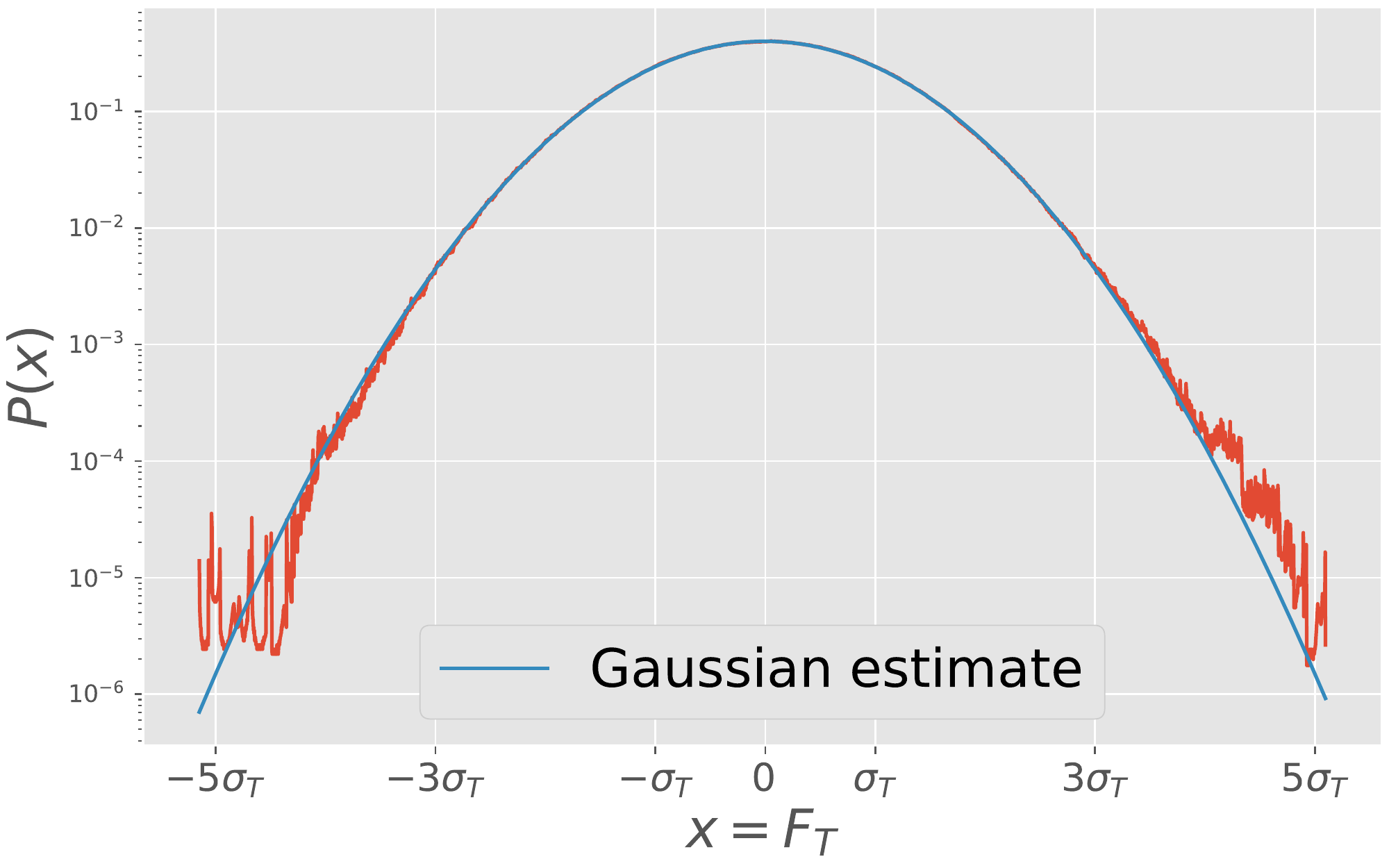}
	\caption{\ac{pdf} of the time-averaged drag $F_T$ with $T = 10\tau_c$. }
	\label{fig:PDF_AVG}
\end{figure}

We discussed previously the phenomenology of extreme fluctuations of the \emph{instantaneous} drag, and identified the sweeping time of the flow past the obstacle as the characteristic lifetime of these events. 
In applications, this duration may be much smaller than the response time of the material structure subject to these fluctuations, justifying a practical interest in the averaged (in time) drag force. 
%
% definition
%
Therefore, another relevant observable is the \textit{time-averaged} drag defined as
\begin{equation}
\label{eq:def_time_averaged_drag}
F_T(t) = \frac{1}{T}\int_t^{t+T} f_d(t) \mathrm{d}t,
\end{equation}
where $f_d(t)$ denotes the instantaneous drag and $T$ is the investigated timescale (response time).
%\sout{Relevant values of $T$ are application dependant.}
In the following, we shall consider $T=10\tau_c$, where $\tau_c$ is the correlation time of the instantaneous drag.
In that case, the PDF of $F_T$ is found nearly Gaussian as a consequence of the Central Limit theorem (see Fig.~\ref{fig:PDF_AVG}).

During a time interval $[t;t+T]$, a fluctuation of $F_T(t)$ may be roughly viewed as the overall contribution of  $T / \tau_c$ independent fluctuations of the {instantaneous} drag $f_d$.  
% phenomenology
%
It is thus legitimate to ask whether a large value of the averaged drag results from a single outstanding fluctuation of the instantaneous drag (case (1)),  or from  an unusual succession of moderate positive fluctuations (case (2)). 
In the same way as in section \ref{sec:extreme_extraction}, one can identify extreme fluctuations of $F_T$ exceeding some fixed threshold $a$, and sample a set of extreme events.
Setting $a=5.2\sigma_T$ with $\sigma_T$ being the standard deviation of $F_T$, $84$ independent extreme events were sampled. Again, this choice results from a compromise between the need to consider large deviations from the mean value and the requirement to sample a sufficient number of events for meaningful statistics. As a rule of thumb, the threshold has been set so that about one hundred events are sampled.

% neither case (1) nor case (2)
%
Fig.~\ref{fig:extreme_avg} displays the time-series $\{f_d(t)\}_{t^{\star} \leq t \leq t^{\star}+T}$ for several extreme fluctuations of $F_T$ occurring at $t=t^\star$.
Qualitatively, it is found that extreme fluctuations of the time-averaged drag can neither be reduced to case (1) nor case (2). 
Indeed, both cases are featured in Fig.~\ref{fig:extreme_avg};
very large value of the averaged-drag appear to result from either a very large fluctuation, or a significant succession of moderate (positive) fluctuations of the instantaneous drag.

\section{Rare-event sampling algorithms}
\label{sec:rare_events_algorithms}

% limitation of direct sampling
%
In the limit of very rare events and complex dynamics such as turbulent flows, the computational cost of  direct sampling becomes prohibitive. Indeed, the return time of fluctuations $f_d \geq a$ scales as $r(a) \propto e^{\ell a}~(\ell>0)$ according to Eq.~\eqref{eq:return_time}.
As a consequence, the computational cost required to sample events of amplitude $f_d \geq a$ through brute-force sampling diverges typically as $e^{\ell a}$.
%% objective
The aim of rare-event sampling algorithms is therefore to sample extreme fluctuations for a  computational cost much lower than their return time. 
Rare-event (sampling) algorithms compute an ensemble of $N$ trajectories $\{\mathbf{x}_n(t)\}_{0\leq t \leq T_a}$ with $n=1 \cdots N$, where $\{\mathbf{x}_n(t)\}_{0\leq t \leq T_a}$ refers to a trajectory of duration $T_a$ in the phase space of the system.
At each step of the algorithm, some trajectories are discarded and others are replicated in order to preferably sample trajectories with extreme fluctuations. The algorithm tracks the ratio of the probability of the new ensemble with the probability of the previous one, allowing the estimation of the statistical bias and therefore the inference of the statistics of extreme events. 

Different algorithms use different \emph{selection rules}. 
The success of the algorithm essentially depends on the quality of the selection rule to detect the precursors of the extreme event \citep{rolland_statistical_2015}. 
However, with the exception of rare analytical cases, the optimal selection rules are not a priori known \citep{lestang_computing_2018}. 
In the following we consider the \acl{ams} (\ac{ams}) and \acl{gktl} (\ac{gktl}) algorithms. They both proved to be efficient for various dynamics but adopt opposite strategies: the \ac{ams} algorithm uses as a selection rule the value of a predetermined score function, whereas the \ac{gktl} uses a selection rule based on the increment of a score function. A complete description of these two algorithms and their operating principles are provided in appendices \ref{app:ams-short} and \ref{app:gktl_description}.

The \ac{ams} algorithm \citep{cerou_adaptive_2007} builds on previous ideas about splitting approaches \citep{KahnHarris1951,glasserman_multilevel_1999}.  
In recent years, it has allowed for the computation of rare events in problems involving a large number of degrees of freedom such as molecular dynamics simulations \citep{aristoff_adaptive_2015,teo_adaptive_2016}
or stochastic partial differential equations for the computation of rare trajectories in the Allen-Cahn equations \citep{rolland_computing_2016}. 
 More recently it has been applied to rare events in stochastic models of wall-turbulence \citep{rolland_extremely_2018} and atmospheric dynamics \citep{bouchet2019rare}. A review of the \ac{ams} algorithm, its history and applications is also available in \citep{cerou2019adaptive}. 

During the last decade, the main theoretical framework for the study of rare events in statistical physics has been the theory of large deviations \citep{touchette_large_2009}.
Alongside, numerical methods have been developed to sample rare events \citep{DelMoralBook}. 
Among them the \ac{gktl} algorithm \citep{giardina_direct_2006} is particularly suitable for estimating the probability of observables which are temporal integrals over a very long period \citep{giardina_simulating_2011,Laffargue_2013}.
Recently the \ac{gktl} algorithm has proven to be extremely efficient to simulate extreme heat waves in a comprehensive climate model \citep{ragone_computation_2018}, with a gain of a factor 100 to 1000 for the computation time.
This achievement already represents a significant leap in the applicability of rare-event sampling to complex dynamical systems. 
The aim of this part of our work is to test rare-event algorithms for fluid-structure interaction in a turbulent flow, which is an unexplored area of application. 

\subsection{Extreme instantaneous drag forces with the \acl{ams} algorithm}
\label{sec:ams}

The \ac{ams} algorithm is here used with trajectories of fixed duration $T_a$~\citep{lestang_computing_2018}.
%
% Quick description of TAMS
%
At each iteration, trajectories with the lowest maxima of the score function $\xi (\mathbf{x}(t),t)$ during the time interval $0\leq t \leq T_a$ are discarded. 
These trajectories are re-sampled by branching them from the remaining trajectories.
The operating principle of the \ac{ams} algorithm is detailed and sketched in Appendix \ref{app:ams-short}.
The objective is here to sample flow evolutions which exhibit extreme fluctuations of the drag $f_d(t)$ acting on the square obstacle.
The observable itself is used as the score function, \emph{i.e.} $\xi (\mathbf{x}(t),t) = f_d(t)$.  
Since the Navier--Stokes dynamics is deterministic, small random perturbations are artificially introduced at branching points; this procedure is detailed in Appendix \ref{app:perturb_branching_time}.
The chaotic dynamics then ensures that re-sampled trajectories separate from their parents over a time interval $\tau_L$, usually referred to as the Lyapunov timescale.
Based on linear response theory, we expect this small perturbation to have a negligible impact on the statistics of the sampled rare events. This has been tested by performing a long simulation of the dynamics and by regularly perturbing the flow. We checked that the obtained statistics of the drag were consistent, within an error of the same amplitude as the perturbation amplitude, with the statistics computed from the (unperturbed) control simulation (see Appendix \ref{app:perturb_branching_time}).

In addition to the score function, two important parameters are the number of trajectories $N$ and their duration $T_a$.
  The size of the ensemble $N$ governs the statistical error affecting quantities averaged over the sampled set of trajectories. Therefore, $N$ should be taken as large as possible to reduce these errors with, nevertheless, a practical limit given by the available computational resources. In this work, we have performed two numerical simulations with $N=32$ and $N=256$.
The duration $T_a$ should be much larger than $\tau_c$~\citep{lestang_computing_2018} but, again, kept small enough to limit the computational cost. In practice, the duration of the trajectories was eventually set to $T_a=5\tau_c$ in both simulations; we checked in particular that larger values of $T_a$ did not improve the results.

\begin{figure}
	\centering
	\includegraphics[width=.7\linewidth]{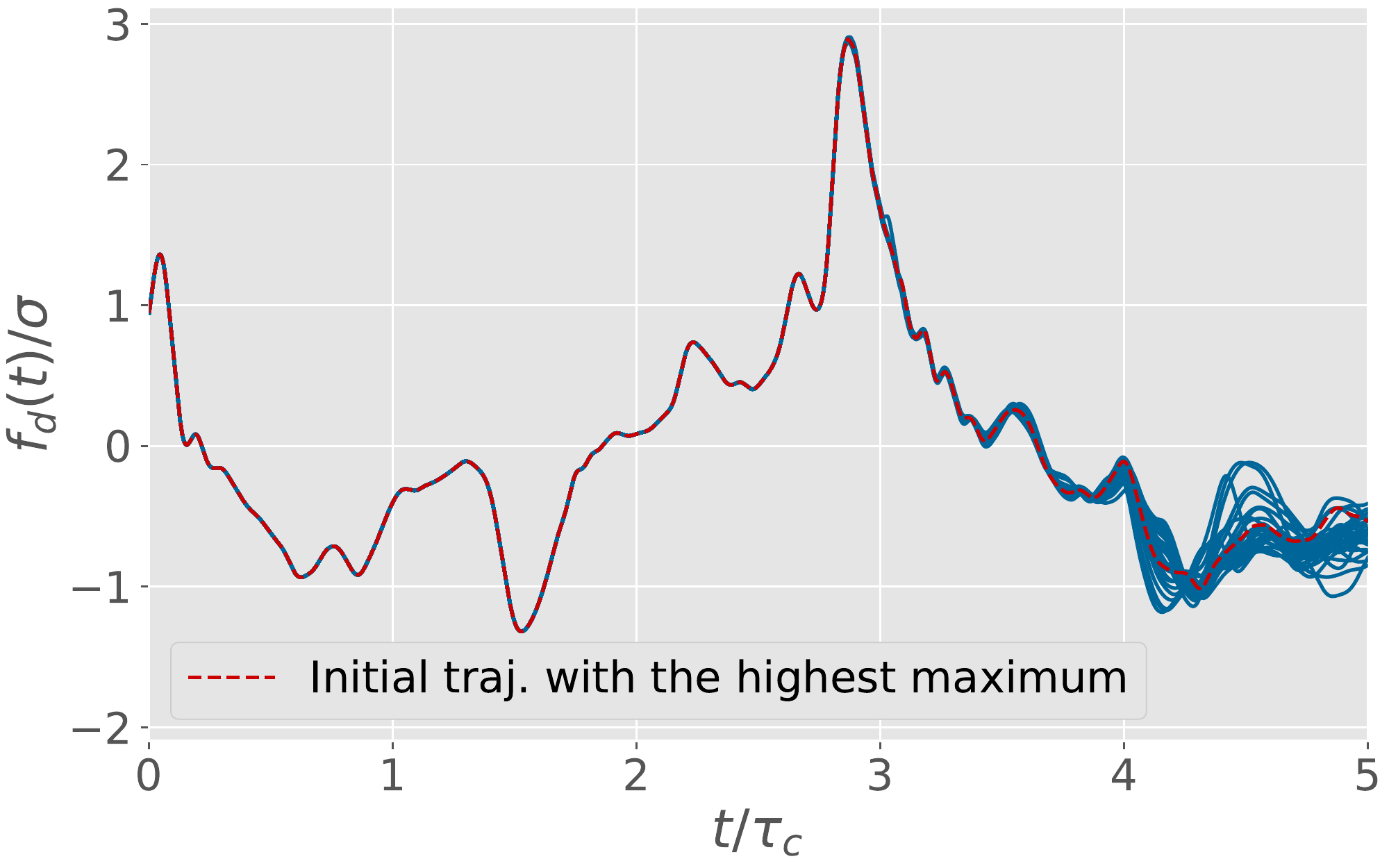}
	\caption{\label{fig:AMS_drag_trajectories} Ensemble of $N = 32$ trajectories after $181$ iterations of the \ac{ams} algorithm. In this experiment, the algorithm is used with the instantaneous drag $f_d(t)$ as score function. Each trajectory has a duration $T_a = 5\tau_c$  where $\tau_c$ is the correlation time of the instantaneous drag.}
\end{figure}

\begin{figure}
	\centering
	\includegraphics[width=.7\linewidth]{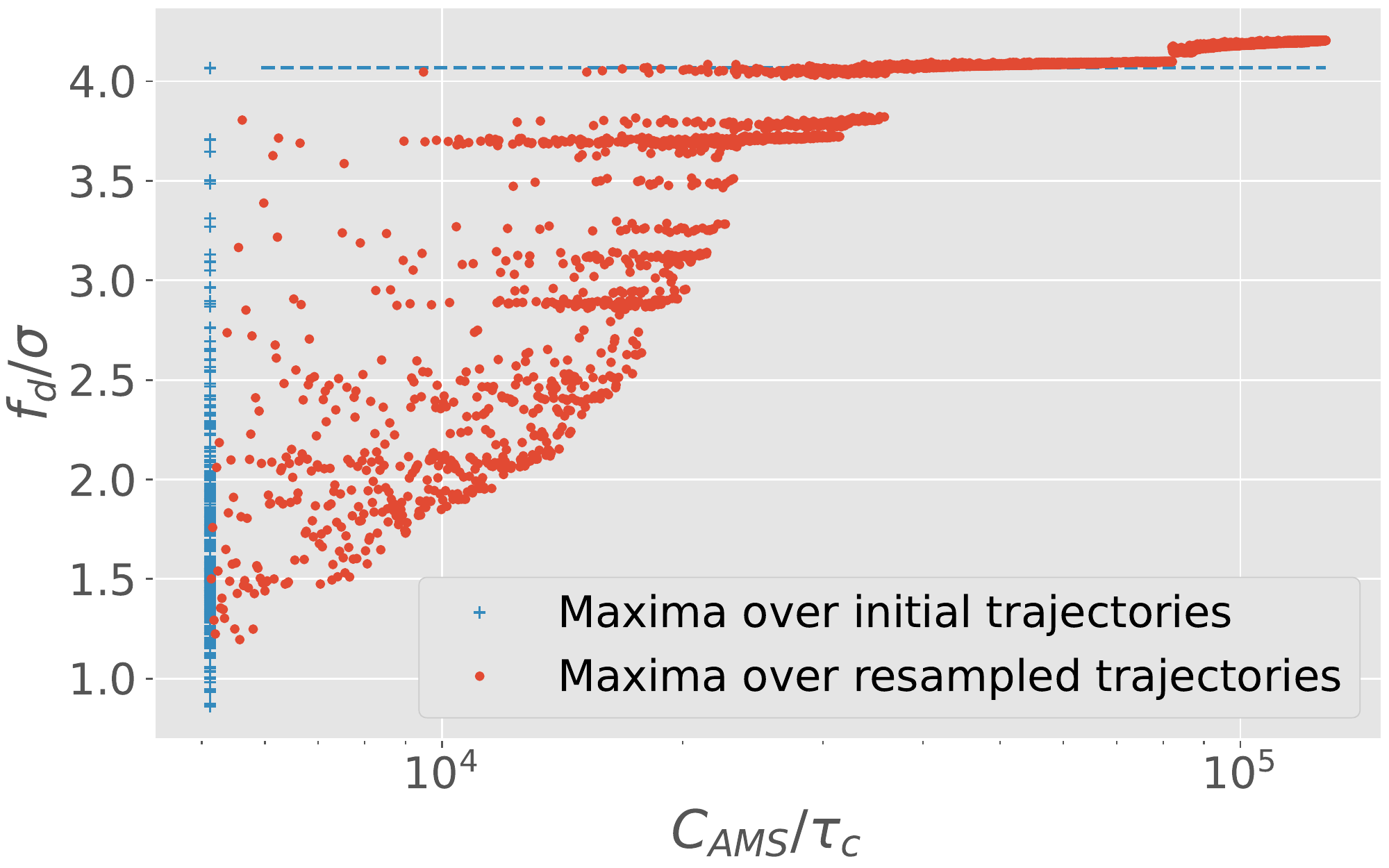}
	\caption{\label{fig:AMS_drag_resampling} Maxima of the instantaneous drag throughout $256$ re-sampled trajectories (vertical axis) as a function of the corresponding computational cost $C_{AMS}$ in correlation-time unit. The \ac{ams} algorithm fails to efficiently sample rare trajectories associated to drag fluctuations higher than the largest fluctuation already captured in the initial ensemble.}
\end{figure}

%pitfall
%
Fig.~\ref{fig:AMS_drag_trajectories} displays the ensemble of drag values after many iterations of the \ac{ams} algorithm with $N=32$. 
Interestingly, one obtains that all trajectories are eventually re-sampled from the same trajectory displaying the highest maximum in the initial ensemble, and overlap over most of their duration.
The algorithm thus fails to enhance the sampling of extreme events. 
This result is also confirmed by increasing the number of initial trajectories to $N=256$.
The maximum drag achieved by the re-sampled trajectories is displayed as a function of the computational cost in Fig.~\ref{fig:AMS_drag_resampling};  
the distribution of the maximal drag for the {initial} trajectories is also shown.
We observe that the trajectories with the lowest maxima of the score function are discarded after a few iterations, and new trajectories with higher maxima are re-sampled. 
However, the re-sampled trajectories never exceed the amplitude of the highest maximum already attained in the initial set of trajectories. A phenomenological explanation is developed in the next paragraph.

It takes a time $\tau_L$ (Lyapunov timescale) before a re-sampled trajectory separates from its parent. In our situation, this ``memory effect'' is related to the fact that the score function is of dimension much smaller (one) than the dimension of the phase space or, in other words, that the score function results from the contribution of a very large number of degrees of freedom.  
As shown in section \ref{sec:dynamical_aspects}, extreme drag fluctuations have a lifetime $\tau_c$ related to the timescale over which a vortex remains trapped against the base of the obstacle. After $\tau_c$, the vortex is swept away by the flow and further large fluctuations of the drag can only result from the trapping of new vortices.
Fig.~\ref{fig:AMS_drag_trajectories} shows that $\tau_c$ is shorter than the Lyapunov's timescale $\tau_L$. 
Therefore, the re-sampling of a trajectory branched close to $t=t^{\star}$ (when the maximum drag occurs) cannot lead to larger values at $t^{\star} \leq t \leq t^{\star}+\tau_L$.
For $t - t^{\star} >\tau_L$, the drag process has lost the memory of the drag fluctuations on which the re-sampling was based and, thus, the probability of observing a new extreme fluctuation is also very low. 

The difference between the typical duration of drag fluctuations $\tau_c$ and the Lyapunov timescale $\tau_L$ may be heuristically associated with the so-called \emph{turbulence rate} \citep{frisch_book}.
As discussed previously, the duration of extreme fluctuations of the drag is closely related to the sweeping time of the flow past the obstacle, and consequently to the mean-flow velocity $U$. On the contrary, the Lyapunov timescale is rather associated with the intrinsic evolution of turbulent fluctuations in the reference frame of the mean flow, \textit{i.e.} with the root mean square velocity $u_{rms}$. The ratio $u_{rms}/U$ (turbulence rate) is much lower than one in our case of grid-generated turbulence, which thus justifies that $\tau_L > \tau_c$.

% perspectives
%
In summary, a straightforward application of the \ac{ams} algorithm with the score function being the drag itself does not allow us to efficiently sample extreme fluctuations.
This behaviour is independent of the choice of $N$ and $T_a$. Increasing the size of the initial ensemble, or the duration of each trajectory, can only increase the amplitude of the global maximum reached initially but does not solve the issue of overlapping trajectories.

\subsection{Extreme time-averaged drag forces with the \acl{gktl} algorithm}
\label{sec:gktl}
The sampling of extreme fluctuations of the time-averaged drag $F_T$ is now examined.
The \ac{ams} algorithm could be used in the same way as before by taking the time-averaged observable itself as the score function.
However, this would lead to similar unsatisfactory results.
For a time-averaged observable, an alternative approach is provided by the \acf{gktl} algorithm \citep{giardina_direct_2006,tailleur_probing_2007,giardina_simulating_2011}.
Similar to the \ac{ams} algorithm, the \ac{gktl} algorithm relies on the simulation of an ensemble of trajectories.
At regular time intervals, some elements of the ensemble are killed and others are cloned according to a weight that depends on the history of the element itself.
The weights are chosen so that, after several iterations of the algorithm, the trajectories in the ensemble are distributed according to a biased probability distribution that favors trajectories related to large values of the time average of the observable.
The \ac{gktl} algorithm belongs to a family of algorithms known as ``go-with-the-winners'' \citep{aldous1994go,grassberger2002go}.
{Similar ideas have already been applied in a wide range of fields such as polymer physics~\citep{grassberger1998perm}, out of equilibrium statistical physics~\citep{PhysRevLett.118.115702}, computer science~\citep{aldous1994go}, dynamical systems~\citep{tailleur_probing_2007}, quantum mechanics~\citep{intro_DMC_kosztin}.}
The application of a go-with-the-winners approach to the computation of large deviations in non-equilibrium systems has first been proposed in \citep{giardina_direct_2006}.
Over the last ten years, it has been successfully applied to investigate rare events in both stochastic \citep{giardina_direct_2006,lecomte_numerical_2007,garrahan2007dynamical} and deterministic systems \citep{giardina_direct_2006,tailleur_probing_2007}.
The operating principle of the \ac{gktl} algorithm is developed in Appendix \ref{app:gktl_description}.

\begin{figure}
	\centering
	\includegraphics[width=0.7\linewidth]{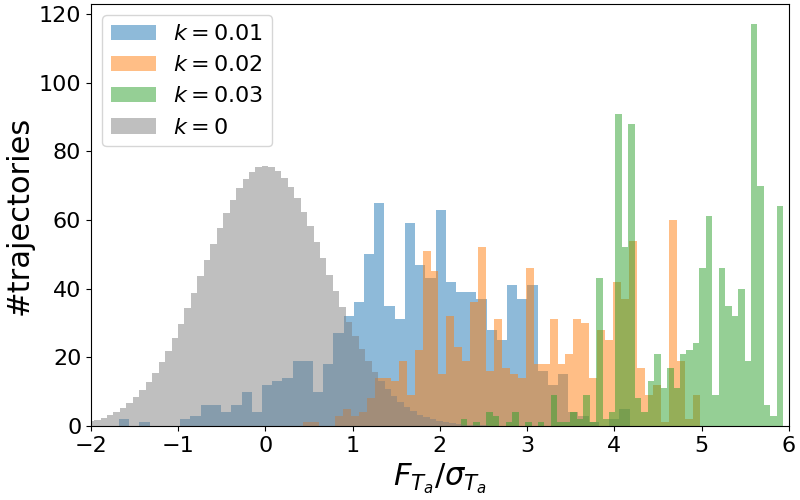}
	\caption{\label{fig:IS_GKTL} Rare-event sampling of the (zero-mean) time-averaged drag $\tilde F_T = F_T - \overline{F_T}$ with $T=10\tau_c$; $\tau_c$ is the correlation time of the instantaneous drag. The shaded \ac{pdf}s are estimated from the biased ensemble resulting from the \ac{gktl} algorithm applied to $N=1024$ trajectories of duration $T_a=T$ with the cloning period $\tau=\tau_c/2$.
		The dashed line refers to the unbiased \ac{pdf} of $\tilde F_T$, \textit{i.e.} obtained from direct sampling or with $k=0$ (no bias) in the \ac{gktl} algorithm.
	}
\end{figure}

The application of the \ac{gktl} algorithm is considered for the flow dynamics introduced in section~\ref{sec:test_flow}. The purpose is to speed-up the sampling of trajectories with extreme fluctuations of the \textit{time-averaged drag}, $F_T$. 
Our observable of interest is therefore the drag $f_d(t)$ and the duration $T_a$ of each trajectory corresponds to the period of averaging, \emph{i.e.} $T_a=T$.    
In a nutshell, trajectories are first evolved in time  independently up to $t=\tau $, with $\tau < T_a$ referring to a cloning period. Then, the selection (and cloning) rules apply according to the average of the observable of interest, here $f_d(t)$, over the interval $[0,\tau]$. 
This procedure is repeated $n-1$ times over the intervals $[\tau, 2\tau],~...,~[(n-1)\tau, n\tau = T_a]$. As a result, the resampled trajectories are distributed according to a probability distribution that is tilted towards large values of the averaged observable. Further details are provided in Appendix~\ref{app:gktl_description}.

%
% Choice of parameters and perturbation
The algorithm depends on three parameters: the cloning strength $k$, the number of trajectories $N$ and the resampling period $\tau$. The higher $k$, the larger is the bias involved in the statistical resampling. 
Similar to the \ac{ams} algorithm, $N$ governs the error affecting the averaged quantities evaluated from the biased ensemble of trajectories, and should be chosen as high as possible. The cloning period $\tau$ determines how often the resampling is performed. 
A small cloning period can result in a loss of information if the clones do not separate from their parents between two cloning steps. On the contrary, choosing $\tau$ much higher than  $\tau_c$  results in insufficient cloning steps.
As a result, a rule of thumb is $\tau \approx \tau_c$.

\begin{figure}
	\centering
	\includegraphics[width=\linewidth]{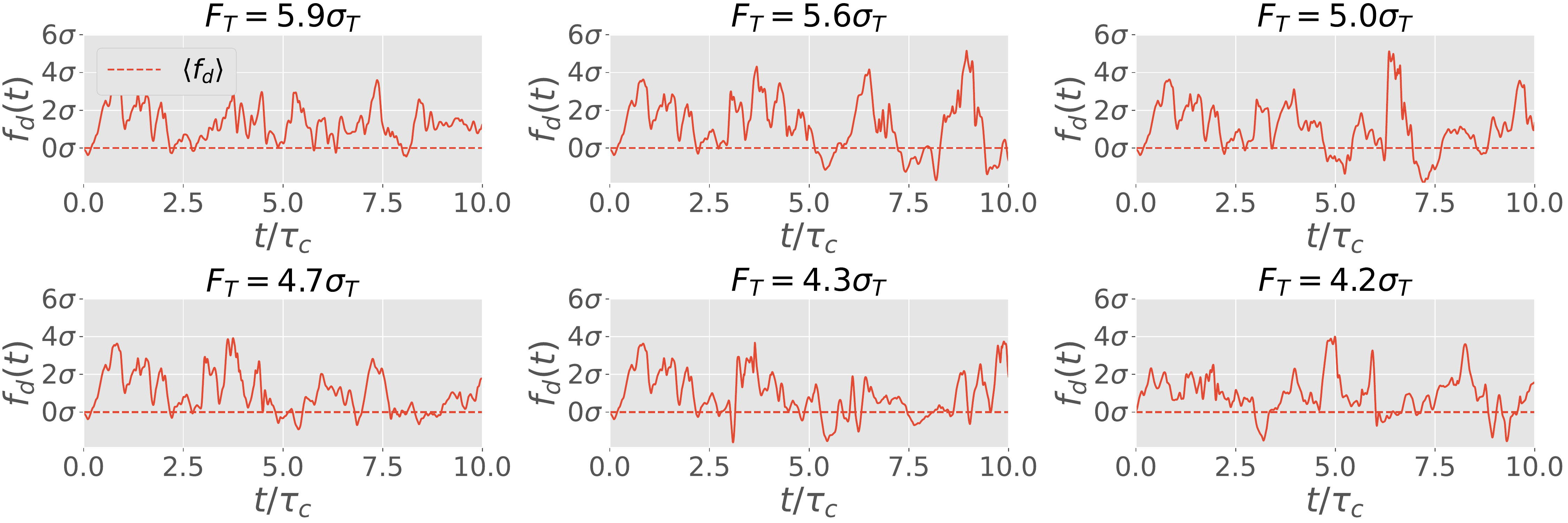}
	\caption{Drag timeseries corresponding to the highest fluctuations of the averaged drag in the ensemble of trajectories sampled  by the \ac{gktl} algorithm. The algorithm was applied with $N = 1024$, $\tau = \tau_c / 2$ and $k = 0.03$. }
	\label{fig:timeseries_extrms_AVG_GKTL}
\end{figure}

\begin{figure}
	\centering
	\includegraphics[width=\linewidth]{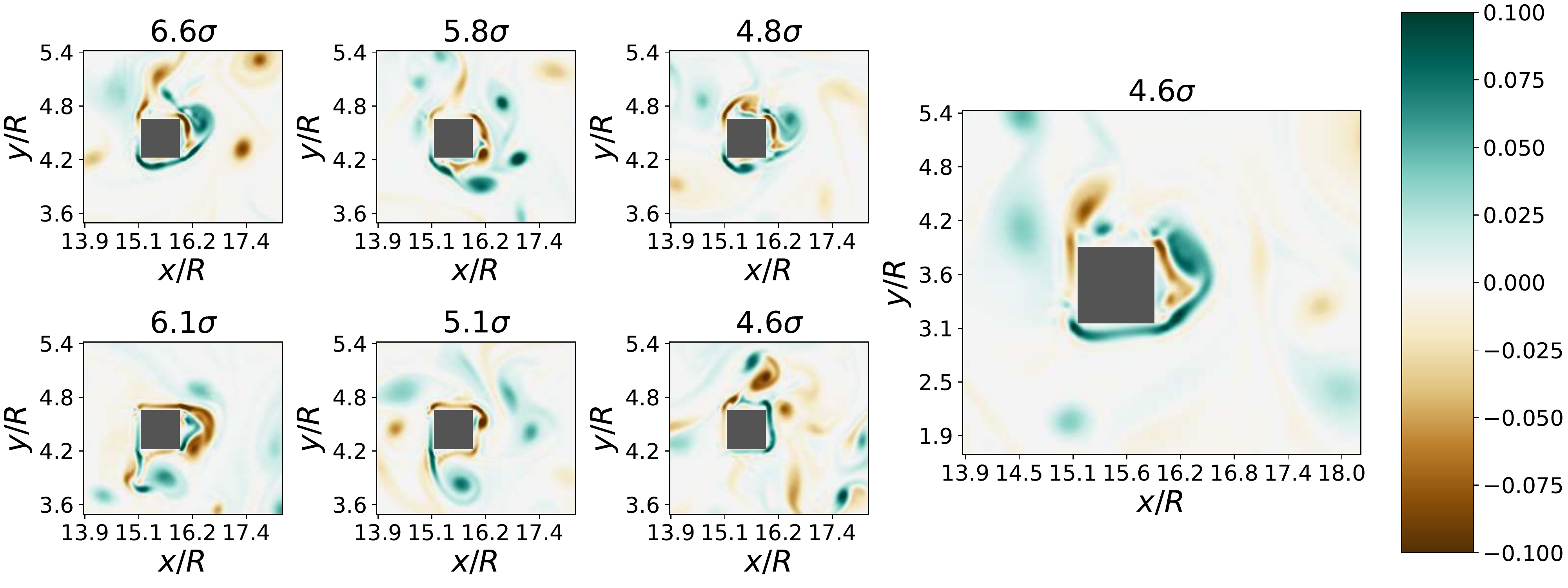}
	\caption{Vorticity field associated with the maximum of the instantaneous drag $f_d$ in  trajectories sampled by the \ac{gktl} algorithm.}
	\label{fig:illustr_extrms_vorticity_GKTL}
\end{figure}

In the following experiments, $N=1024$ and $T_a = 10\tau_c$, which yields a computational cost $C_{gktl} = N \times T_a \approx 10^4 \tau_c$. We fixed $\tau = \tau_c /2$ as a satisfactory compromise in order to ensure an efficient sampling \citep{lestang:tel-01974316}.
Three numerical experiments corresponding to three different values of the strength parameter $k$ have been carried out.
Fig.~\ref{fig:IS_GKTL} shows the histograms of the (zero-mean) time-averaged drag for the three ensembles of trajectories, in addition to the unbiased histogram based on a gaussian approximation of the \ac{pdf} of the time-averaged drag.
As expected, the algorithm samples preferentially trajectories with a higher value of the averaged drag. Furthermore, the higher $k$, the stronger the bias.
%
%
% limitations
%
Nevertheless, it should be mentioned that for a given number $N$ of trajectories, there is necessarily an upper limit $k_{max}$ of the strength parameter over which the finite number of trajectories becomes detrimental to the efficiency and accuracy of the selection procedure.
For $k \gtrsim k_{max}$, the re-sampling relies only on a small number of ``independent trajectories'' and most of the trajectories in the biased ensemble overlap. 
This effect is highlighted in Fig.~\ref{fig:IS_GKTL} where the histogram corresponding to $k=0.03$ becomes artificially peaked.
In the present simulations with $N=1024$ trajectories, one can empirically estimate that  $k$ should be kept smaller than $k_{max} \approx 0.03$ to ensure the independence of the trajectories in the biased ensemble.
Fig.~\ref{fig:timeseries_extrms_AVG_GKTL} displays the drag signal for the extreme trajectories sampled by the \ac{gktl} algorithm.
In addition, Fig.~\ref{fig:illustr_extrms_vorticity_GKTL} displays the vorticity field related to the maximum of the drag in sampled trajectories.
One observes that the extreme events are consistent with the picture of strong vorticity being trapped in the vicinity of the base of the obstacle, as pointed out in section~\ref{sec:instantaneous_drag}. Qualitatively, the related drag signals are also found very similar to those obtained from direct sampling. 
However, it should be stressed that the computational cost for sampling the events shown in Fig.~\ref{fig:timeseries_extrms_AVG_GKTL} and Fig.~\ref{fig:illustr_extrms_vorticity_GKTL} is roughly one hundred time lower than the computational cost required to capture by direct sampling the events displayed in  Fig.~\ref{fig:top_4_events_vorticity} and Fig.~\ref{fig:extreme_avg}.

\subsubsection{Computation of return times}
\label{sec:return_times}

Importantly, the \ac{gktl} algorithm provides an ensemble of trajectories over which statistics of rare events can be evaluated. In this section, we show in particular that this ensemble allows for the estimate of return times for fluctuation amplitudes that would be unreachable by direct sampling (with the same computational cost). The method for estimating return times is described in Appendix~\ref{sec:return-time-gktl}.

Fig.~\ref{fig:return_times_gktl} displays the return times for extreme fluctuations of the time-averaged drag obtained by using the \ac{gktl} algorithm with two different values of the strength parameter $k$.  
Both estimates have been obtained with the same computational cost $N\times T_a$.
An estimate given by direct sampling with a time-series of the same effective duration $T_{tot}=N\times T_a$ is also shown. 
Whilst a direct approach cannot access (by definition) events with a return time greater than $T_{tot}$, the \ac{gktl} algorithm allows us to estimate the statistics of time-averaged drag fluctuations having a return time several orders of magnitude above $T_{tot}$. Alternatively, for a fixed target return time, the use of the \ac{gktl} algorithm can reduce the computational cost of the estimation by several orders of magnitude. This is  obviously a major advantage of this rare-event sampling algorithm. 
\begin{figure}
	\centering
	\includegraphics[width=.7\linewidth]{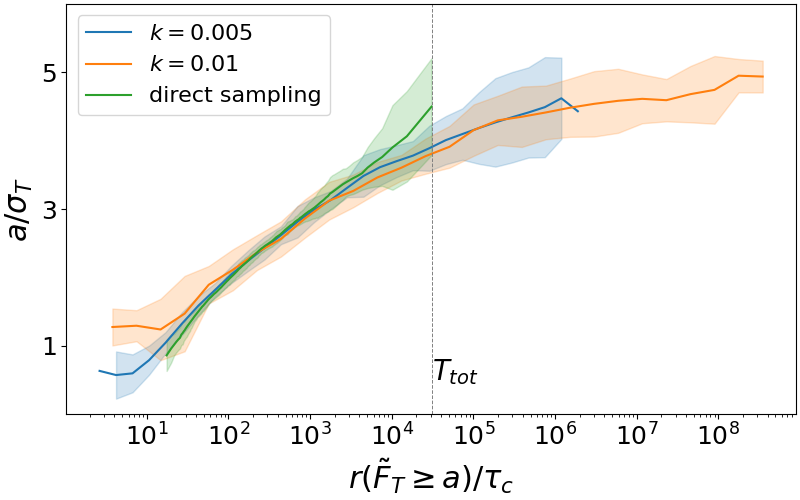}
	\caption{\label{fig:return_times_gktl} Return times for the time-averaged drag acting on the square obstacle. $\tilde{F}_T$ denotes the time-averaged drag with zero mean. The blue and red lines are obtained from the biased ensemble of trajectories generated by the \ac{gktl} algorithm with $N=1024$, $T_a=30\tau_c$ and a cloning period $\tau=\tau_c/2$. The green line is the return times obtained from a single timeseries of duration equal to the computational cost of both \ac{gktl} experiments. Uncertainty ranges for the \ac{gktl} estimates are computed as the standard deviation over a set of 10 independent experiments. Uncertainty ranges for the direct estimation are computed as the standard deviation over a ensemble of direct estimates resulting from 60 independent timeseries.}
\end{figure}
\section{Conclusion}
\label{conlusion}
The dynamics and statistics of extreme fluctuations of the drag acting on a squared obstacle mounted in a turbulent channel flow (in two dimensions) have been investigated numerically through both direct and rare-event sampling methods. 
By means of direct sampling based on a very long simulation, it was first observed that such extreme events are generically related to the trapping of a strong vortex against the base of the obstacle by a streamline cap.
This arrangement does not persist over time, however, since the main flow eventually sweeps away the surrounding fluid structures responsible for this trapping. Therefore, the lifetime of extreme drag fluctuations is found of the order of the sweeping time of the flow past the obstacle, and the corresponding drag signal is very peaked around extreme fluctuations.
In addition, it was found that extreme fluctuations of the time-averaged drag do not preferentially result from a small number of very large fluctuations or an exceptional succession of moderate fluctuations that pile up to yield a large value of the average; both configurations are observed. 
A second part of this study has been dedicated to the application of two representative rare-event algorithms, namely the \acl{ams} and the \acl{gktl} algorithms, to our fluid-mechanical problem. These algorithms rely on selection rules that determine how an initial ensemble of trajectories is evolved to possibly enhance the sampling of extreme fluctuations. 
On the one hand, the \ac{ams} algorithm fails (for this specific application) to generate trajectories exhibiting extreme events at a better rate than a direct sampling. This result can be related to the phenomenology of extreme drag fluctuations, whose lifetime is shorter than the timescale over which the replicated trajectories manage to separate from their duplicate.
Therefore, the algorithm is unable to benefit from the precursors of extreme fluctuations to enhance the realization of new rare-event trajectories. 
On the other hand, the \ac{gktl} algorithm leads to a computational gain of several orders of magnitude. The latter is based on the cumulative evolution of the system rather than its instantaneous behaviour, as opposed to the \ac{ams} algorithm. The downside is that the algorithm provides only statistics of extreme values of the time-averaged fluctuations. This can nevertheless be of great interest for computing return times of extreme (time-averaged) fluctuations, for instance.   

Selection rules at the heart of rare-event algorithms rely heavily on the choice of a score function that drives the selection and replication of trajectories. 
In this study, the observable itself (the drag) has been chosen as the score function.
Optimizing the choice of this score function is desirable (especially for the \ac{ams} algorithm) but difficult in practice, since it should account for the phenomenology of the extreme events themselves, for example how these events build up. Our study shows this quite clearly. A possible direction would then be to take advantage of recent advances in \emph{learning methods} to dynamically optimise the score function.

\section{Acknowledgements}
The authors thank Francesco Ragone, Corentin Herbert, Charles-Edouard Bréhier and Eric Simonnet for useful discussions and suggestions on various aspects of this work.
T.L and F.B acknowledge support from the European Research Council under the European Union's seventh Framework Program (FP7/2007-2013 Grant Agreement No. 616811).
Simulations have been performed on the local HPC facilities at École Normale Supérieure de Lyon (PSMN) and École Centrale de Lyon (PMCS2I).
The facilities at PSMN are supported by the Auvergne-Rhône-Alpes region (GRANT CPRT07-13 CIRA) and the national Equip@Meso grant (ANR-10-EQPX-29-01).

\appendix
\section{The \acl{lbm}}
\label{app:lbm}

\begin{figure}
	\centering
	\includegraphics[width=0.3\linewidth]{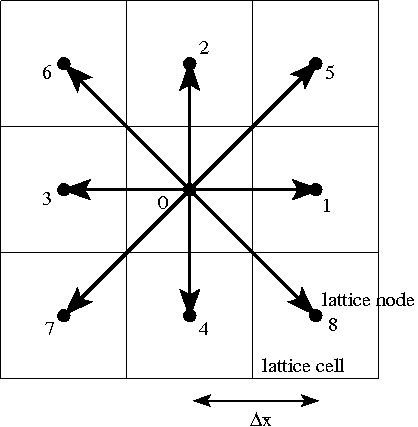}
	\caption{Sketch of the D2Q9 lattice. Particles move exactly from a lattice node towards one of its nine neighbours (including the node itself) during one time step. By definition, the lattice spacing is related to the time step by $\Delta x/ \Delta t = \sqrt{3} c_s$ where $c_s$ is interpreted as a speed of sound.}
	\label{fig:D2Q9}
\end{figure}

% details about LBM
In the LB method, the fluid is viewed as a population of particles that collide, redistribute and propagate along the different links of a discrete lattice (see \citep{kruger_lattice_2017} for a comprehensive introduction). 
In our two-dimensional situation, the so-called D2Q9 lattice with only nine possible velocities $\{\mathbf{c_i}\}_{i=0...8}$ at each node has been adopted (see  Fig.~\ref{fig:D2Q9}).
Locally, the macroscopic flow variables (per unit volume) are recovered by summing over the densities of particles $\{f_i\}_{i=0...8}$ moving with the different velocities, i.e.
\[
\rho(\mathbf{x},t) = \sum_i f_i(\mathbf{x},t) \quad \mathrm{and}\quad \rho(\mathbf{x},t) \mathbf u(\mathbf{x},t) = \sum_i f_i(\mathbf{x},t) \mathbf{c_i}
\]
for the mass density and the fluid momentum respectively. The assumption of weak compressibility (for an ideal gas) is made so that the pressure is directly proportional to the mass density: $p = c_s^2 \rho$ where $c_s$ is interpreted as a speed of sound.  

% algo
%
The complexity of the flow emerges from the repeated application of simple rules of streaming and collision. The \ac{lbm} advances the local densities of particles $f_i(\mathbf{x},t)$ moving with velocities $\mathbf{c}_i$  in a two-step procedure. Namely, an \emph{exact} streaming step 
\[
f_i(\mathbf{x}+\mathbf{c}_i \Delta t, t + \Delta t) = f_i^{\mathrm{out}}(\mathbf{x},t)
\]
during which particles move with their own velocity to a neighbouring node, and an instantaneous collision step
\[
f_i^{\mathrm{out}}(\mathbf{x},t) = -\frac 1 {\tau_\nu} \left(f_i(\mathbf{x},t) - f_i^\mathrm{eq}(\mathbf{x},t) \right)
\]
which achieves a relaxation of local densities towards an absolute equilibrium (at the macroscopic level). The time-scale $\tau_\nu$ (in lattice unit) is related to the kinematic viscosity of the fluid by 
\[
\nu = \left( {\tau_\nu} - \frac 1 2 \right) c_s^2 ~\Delta t
\]
This simplification of the collision kernel is known as the BGK approximation in the kinetic theory of gas \citep{BGK}.
The equilibrium function is given by
\[
f_i^\mathrm{eq}(\mathbf{x},t) = w_i  \rho(\mathbf{x},t) \left( 1 + \frac{\mathrm u(\mathbf{x},t) \cdot \mathbf{c_i}}{c_s^2} +
\frac{u_\alpha(\mathbf{x},t) u_\beta(\mathbf{x},t)({c_i}_\alpha {c_i}_\beta - c_s^2 \delta_{\alpha\beta})}{2 c_s^4} \right)
\] 
with the weight factors $w_0=4/9,~w_{1...4} = 1/9$ and $w_{5...8}=1/36$ for the D2Q9 lattice. 
This discrete Lattice Boltzmann scheme is second-order accurate in $\Delta x $ and compliant to the weakly-compressible Navier-Stokes equations with a third-order error in $\mathrm{Ma}=|\mathbf{u}|/c_s$ as the lattice spacing vanishes, i.e. $\Delta x \to 0$ \citep{succi_book}. 

As mentioned before, the pressure is directly accessible from the mass density: $p = \rho c_s^2$. The viscous stress is also obtained easily from the densities of particles by
\[
\tau^\mathrm{visc.}_{\alpha \beta} = -\frac{\nu}{\tau_\nu ~ c_s^2 \Delta t} \sum_i  {c_i}_\alpha {c_i}_\beta (f_i - f_i^\mathrm{eq})
\]
so that the total stress expresses as
\begin{equation}\label{eq:def_stress}
\tau_{\alpha \beta} = -  c_s^2 \sum_i f_i ~ \delta_{\alpha\beta}  - \frac{\nu}{\tau_\nu ~ c_s^2 \Delta t} \sum_i  {c_i}_\alpha {c_i}_\beta (f_i - f_i^\mathrm{eq})
\end{equation}
Finally, let us mention that in the present context of turbulent flows, the single-relaxation-time BGK collision has been replaced by a multi-relaxation-time procedure based on central moments with an improved stability \citep{De_Rosis_2016}.
\section{The \ac{ams} algorithm}
\label{app:ams-short}

\subsection{Operating principle}

\begin{figure}
	\centering
	\includegraphics[width=\linewidth]{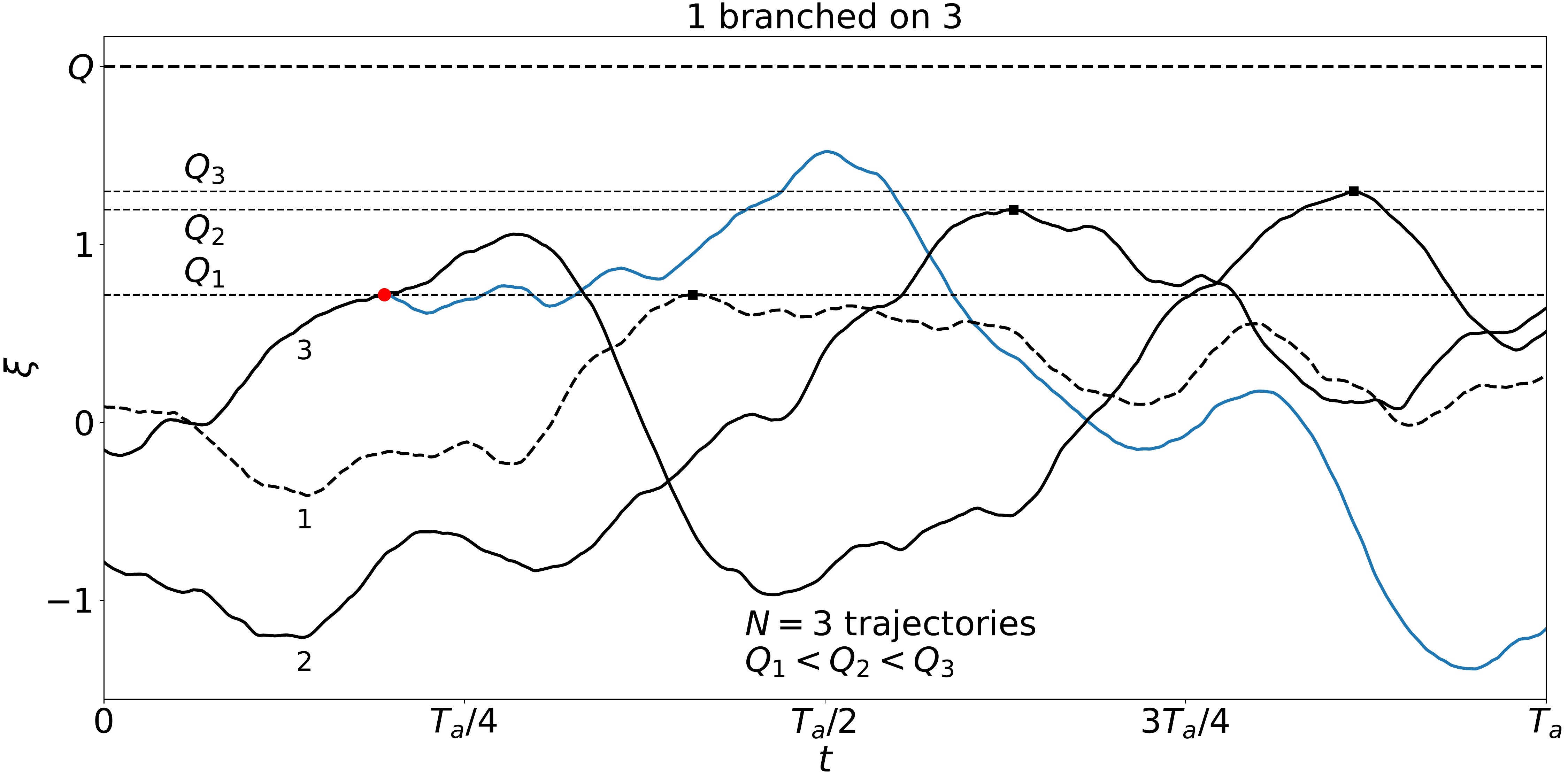}
	\caption{\label{fig:illustr_AMS} Operating principle of the \ac{ams} algorithm with three trajectories of duration $T_a$. $\xi(x(t))$ is a prescribed score function defined along each trajectory. 
	Selection and branching rules are applied in order to reach a maximum score value within the time interval $[0,T_a]$. }
\end{figure}

The operating principe of the AMS algorithm is sketched in Fig.~\ref{fig:illustr_AMS}.
The trajectories 1, 2 and 3 represent the evolution of the \emph{score function} for the current ensemble of trajectories. On the basis of their respective maximum: $Q_1$, $Q_2$ and $Q_3$, the trajectory with the lowest maximum is discarded in the ensemble (dashed line). Among the two remaining trajectories, trajectory 3 is chosen randomly and copied until it reaches the value $Q_1$. It is then simulated from this branching point to the final time $T_a$. In the present case of deterministic dynamics, a small perturbation is introduced at the branching to separate the trajectories. This re-sampling procedure can be iterated $J$ times or until all trajectories do exceed a fixed threshold $Q$.

\subsection{Application to a simple case: the \acl{ou} process}
	
	\begin{figure}
		\centering
		\includegraphics[width=.7\linewidth]{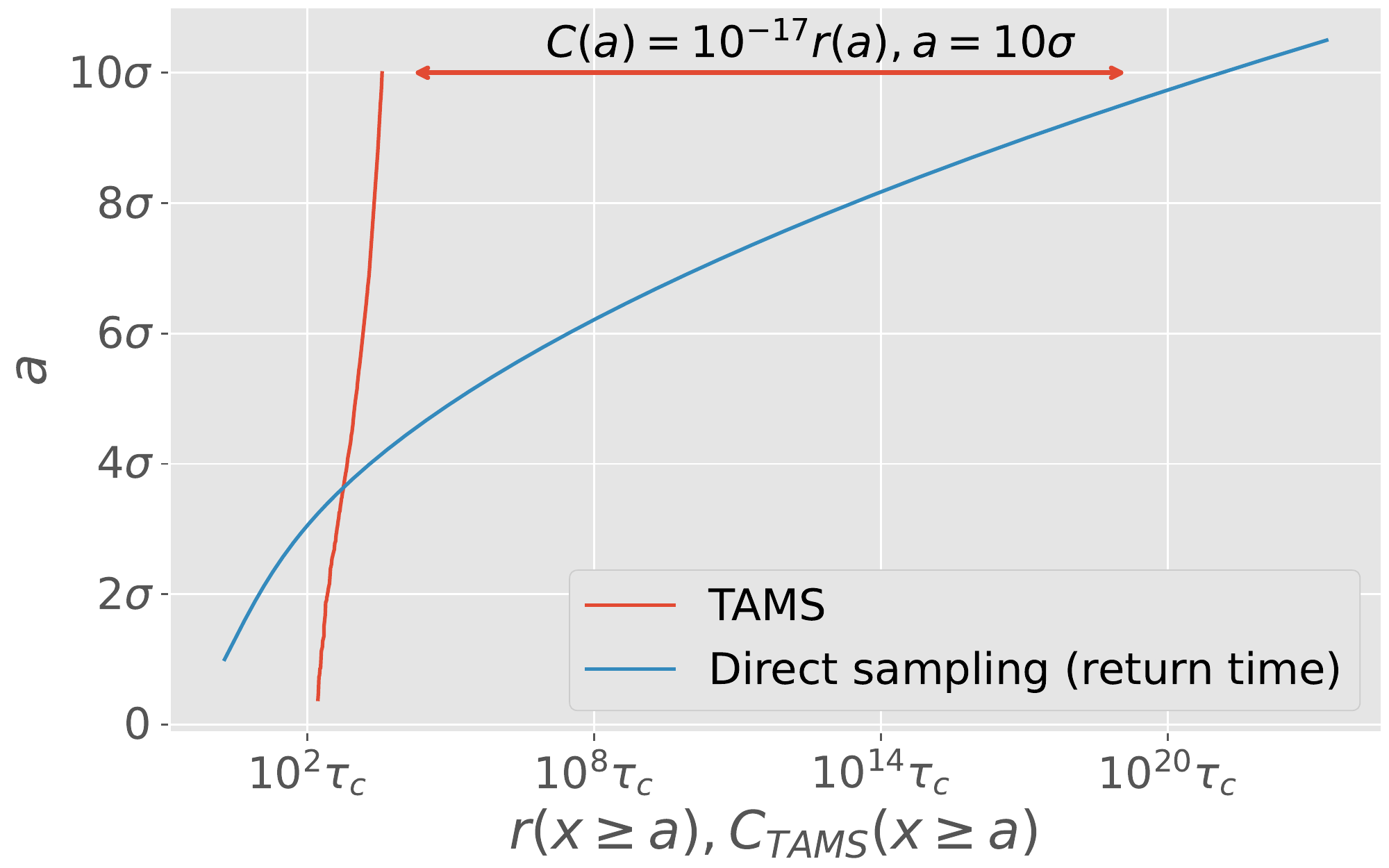}
		\caption{Efficiency of the \ac{ams} algorithm with respect to direct sampling in the case of an Ornstein-Uhlenbeck process \citep{lestang_computing_2018}. The red line represents the evolution of the maximum obtained from re-sampled trajectories as a function of the computational cost $C_{AMS}$. The blue line is the analytical solution for the return time of amplitude $a$.}
		\label{fig:comparaison_temps_de_retour}
	\end{figure}
	A one-dimensional \acl{ou} process is considered:
	\begin{equation}
	\label{eq:ou}
	\dot{x} = -x + \eta (t),
	\end{equation}
	where $\eta$ is a Gaussian noise with $\langle \eta(t)\eta(t-t')\rangle = \delta(t-t')$.

	The \ac{ams} is applied  to a set of $N=32$ trajectories $\{x_n(t)\}_{0\leq t \leq T_a}$ with $T_a=5\tau_c$.
	Let us note that the correlation time is $\tau_c = 1$ for the process defined by Eq.~\eqref{eq:ou}.
	Our objective is to sample fluctuations $x\geq a$ with $a$ being very large compared to the typical values of $x$.
	The score function is simply $x(t)$ and only one trajectory is re-sampled at each iteration.
	The computational cost of the algorithm after $J$ iterations is therefore related to the simulation of the $N$ initial trajectories and the number of re-sampled trajectories.
	Fig.~\ref{fig:comparaison_temps_de_retour} compares the computational cost of the \ac{ams} algorithm with that of a direct sampling. 
	In the latter, the typical computational cost is simply the return time $r(a)$.
	One can see that the successive re-samplings of the \ac{ams} algorithm lead rapidly to trajectories exhibiting extreme fluctuations.
	For large $a$, the computational cost is many orders of magnitude lower than that obtained by direct sampling.
	
	The \acl{ou} process showcases the efficiency of the \ac{ams} algorithm.
	However, the state space is here one-dimensional and the choice of the score function is straightforward: it is $x$ itself.
	In addition, the noise term in Eq.~\eqref{eq:ou} has no correlation in time, which implies that newly generated trajectories quickly separate from their parents. Such favourable features do not persist in the case of fluid dynamics.
\section{The \acl{gktl} algorithm}
\label{app:gktl_description}

\subsection{The operating principle}
\label{sec:gktl-operating-principle}

The \ac{gktl} algorithm is based on the simulation of an ensemble of $N$ trajectories $\left\{\mathbf{x}_{n}(t)\right\}_{0\leq t \leq T_a}$ with $ n =1 \cdots N$ starting from independent random initial conditions.
Let us consider a real-valued observable of interest $A(\mathbf{x}(t))$, {\emph{e.g.} the drag $f_d(t)$}, and introduce a cloning period $\tau$.
At time instants $t_{i}=i\tau$ with $i=1,~2,~...,~T_{a}/\tau$ ($T_{a}$ is a multiple of $\tau$) a weight $W_{n}^{i}$ is assigned to each trajectory. This weight is defined ($t_0=0$) by
\begin{equation}
W_{n}^{i}=\frac{e^{k\intop_{t_{i-1}}^{t_{i}}A(\mathbf{x}_{n}(t))dt}}{R_{i}}\quad \mbox{with the normalisation factor} \quad R_{i}=\frac{1}{N}\sum_{n=1}^{N}e^{k\int_{t_{i-1}}^{t_{i}}A(\mathbf{x}_{n}(t))dt}
\label{eq:Weight}
\end{equation}
so that $\sum_{n=1}^N W_n^i = N$.
{The weights $\{W_{n}^{i}\}_{n=1\cdots N}$ determine how many copies of each trajectory are made at time $t=t_i$. The parameter $k$ characterizes the amplitude of the statistical bias involved in the algorithm (see Fig.~\ref{fig:IS_GKTL}). For more information about the practical implementation of the algorithm, the interested reader can refer to~\citep{brewer2018efficient, lestang:tel-01974316}}.
The application of this re-sampling at each step $t_i$ eventually leads to a biased sampling in the trajectory space; the trajectories corresponding to extreme values of $\int_{0}^{T_a}A(\mathbf{x}_{n}(t))dt$ have a higher probability.
The sampled biased distribution writes
\begin{align}
\mathbb{P}_{k}\left(\left\{ \mathbf{X}(t)\right\} _{0\leq t\leq T_{a}}=\left\{ \mathbf{x}(t)\right\} _{0\leq t\leq T_{a}}\right) &\underset{N\rightarrow\infty}{\sim} \frac{e^{k\int_{0}^{T_{a}}A(\mathbf{x}(t))dt}}{Z(k,T_a)}\mathbb{\mathbb{P}}_{0}\left(\left\{ \mathbf{X}(t)\right\} _{0\leq t\leq T_{a}}=\left\{ \mathbf{x}(t)\right\} _{0\leq t\leq T_{a}}\right),
\label{eq:Biased_Path_Approximation}
\end{align}
where
$\mathbb{P}_{0}\left(\left\{ \mathbf{X}(t)\right\} _{0\leq t\leq T_{a}} = \left\{ \mathbf{x}(t)\right\} _{0\leq t\leq T_{a}}\right)$ 
refers formally to the probability of observing the trajectory 
$\left\{ \mathbf{x}(t)\right\} _{0\leq t\leq T_{a}}$.
The normalisation factor is given by $Z(k,T_a)=\prod_{i=1}^{T_a/\tau}R_i$.
One can mention that
\begin{equation}
\label{eq:mean_field}
Z(k,T_a) \underset{N\to \infty}{\sim} \mathbb{E}_0\left[e^{k\int_{0}^{T_{a}}A(\mathbf{X}(t))dt}\right],
\end{equation}
with $\mathbb{E}_{0}$ being the expectation value with respect to the
distribution $\mathbb{P}_{0}$.
This result relies on the \textit{mean-field approximation}
\begin{equation}
R_{i}=\frac{1}{N}\sum_{n=1}^{N}e^{k\int_{t_{i-1}}^{t_{i}}A(\mathbf{X}_{n}(t))dt}\underset{N\rightarrow\infty}{\sim} Z(k,t_i)= \mathbb{E}_{i}\left[e^{k\int_{t_{i-1}}^{t_{i}}A(\mathbf{X}(t))dt}\right],
\label{eq:Mean_Field_Approximation}
\end{equation}
where $\mathbb{E}_{i}[.]$ denotes the expectation value with respect to the biased distribution $\mathbb{P}_k^{(i)}$ obtained after $i$ cloning steps.
The typical relative error related to this approximation can be shown to be of order $1/\sqrt{N}$ for a family of rare-event algorithms including the \ac{gktl} algorithm~\citep{DelMoralBook,DelMoral2013}.
Rejected trajectories are discarded from the statistics.
Eventually, an effective ensemble of $N$ trajectories of duration $T_{a}$ is obtained, distributed according to $\mathbb{P}_{k}$.

 A key feature of the \ac{gktl} algorithm is that all resampled trajectories are solutions of the original dynamics. 
%This is true for stochastic dynamics.
Nevertheless, it should be noted that a small random perturbation is introduced in the cloning procedure to force clones from the same trajectory to separate, \emph{i.e.} artificial randomness is introduced so that the cloning procedure is effective for deterministic dynamics.   
	As for the \ac{ams} algorithm, it has been checked that this perturbation did not affect the statistics of the sampled trajectories.
%
% 
%However, similarly to the case of \ac{ams}, randomness must be artificially introduced for the cloning procedure to be effective for deterministic dynamics in order to separate the clones. We have proceeded in the same way as as described in section~\ref{sec:ams_drag}.
%
%
Eventually, the sampled trajectories obtained with the \ac{gktl} algorithm can be used to compute the statistical properties of any observable with respect to the distribution $\mathbb{P}_{0}$ from the distribution $\mathbb{P}_{k}$ by using Eq.~\eqref{eq:Biased_Path_Approximation}.

\subsection{Computation of return times}
\label{sec:return-time-gktl}

Each trajectory in the biased ensemble results in a timeseries of the time-averaged drag
\begin{equation}
\label{eq:time_averaged}
F_T^{(j)}(t) = \int_{t-T}^{t}f_d^{(j)}(\tau)d\tau, \quad t\in [T,T_a]  
\end{equation}
and the return time of a fluctuation $F_T \geq a$ is given by~\citep{lestang_computing_2018}
\begin{equation}
r(a) = - \frac{T_a - T}{\ln (1-\mathbb{P}(F_T \geq a))}.
\end{equation}
The probability $\mathbb{P}(F_T \geq a)$ can be estimated from the biased ensemble by inverting Eq.~\eqref{eq:Biased_Path_Approximation}
\begin{equation}
\mathbb{P}(f_d \geq a) \approx \frac{1}{N}\sum_{j=1}^{N}e^{T_a \lambda(k)}e^{k T_a  F_T^{(j)}}s_j(a)
\end{equation}
with $s_j(a) = 1$ if $\max_{T\leq t \leq T_a}[F_T^{(j)}] \geq a$ and $s_j(a) = 0$ otherwise, \emph{i.e.} by summing the weights of the timeseries which maximum is larger than $a$.
\section{Perturbation at branching time}
\label{app:perturb_branching_time}
\begin{figure}
  \centering
\includegraphics[width=.7\linewidth]{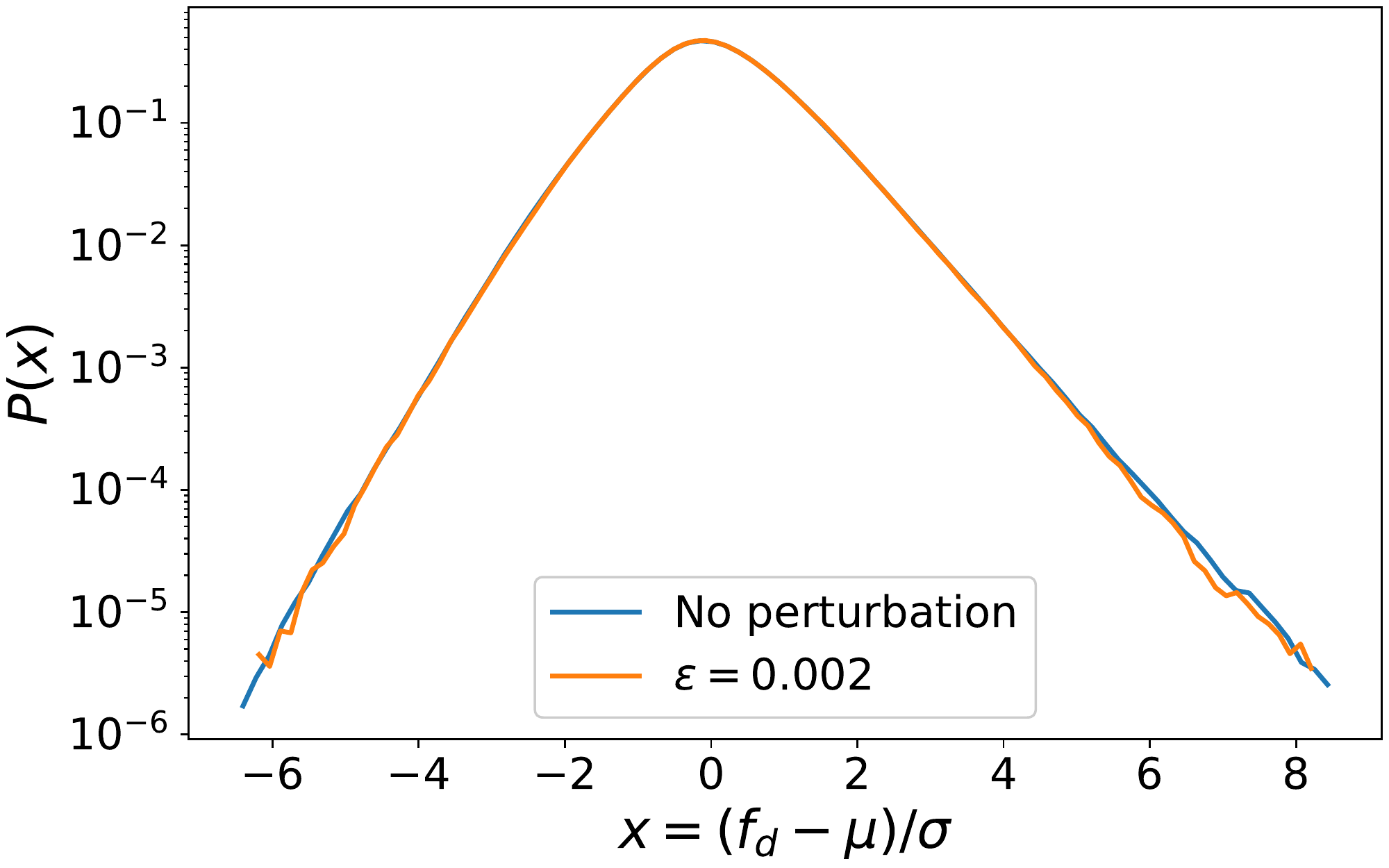}
\caption{\ac{pdf}s of (normalised) drag fluctuations obtained from the perturbed (at branching points) and unperturbed numerical simulations; $\epsilon$ is the relative amplitude of the perturbation.}
\label{fig:pdf_drag_with_perturbation}
\end{figure}

The application  of the \ac{ams} and \ac{gktl} algorithms to a \emph{deterministic} dynamical system requires a perturbation at initial time of the
resampled trajectories. In the absence of such perturbation, the resampling would not yield new trajectories but exact copies of the original trajectories.
In the framework of Lattice Boltzmann simulations, the state of the system is described at a mesoscopic level by the particle densities $\{f_i(\mathbf{x},t)\}_{i=0\cdots8}$ (see appendix~\ref{app:lbm}).
Therefore, the perturbation of the solution at time $t_0$ applies directly to the $f_i$'s with 
\begin{equation}
f_{i}(\mathbf{x},t_0) \longrightarrow f_i(\mathbf{x},t_0) + \epsilon \sum_{n=1}^{N_s} \alpha_{n}f_{i}^{(n)}(\mathbf{x}),
\label{eq:perturb_pop}
\end{equation}
where the $\alpha_n$'s are random numbers uniformly picked in the interval $[0,1]$, $\epsilon$ is the relative amplitude of the perturbation and $\{f_i^{(n)}\}_{n=1 \cdots N_s}$ is a set of arbitrary snapshots of the flow. 
In other words, the perturbation is a  random linear combination of $N_s$ snapshots.
Eventually, the perturbed densities are rescaled so that the mass of the system is preserved.
In practice, we chose $\epsilon = 0.002$ and $N_s = 10$ to ensure that the perturbations remain sufficiently small, random and independent.
In order to check that this perturbation does not impact the statistics of drag fluctuations, a long simulation of duration $T_{tot} = 10^5 \tau_c$ has been performed with a periodic perturbation (with period $\tau_c / 2$) mimicking the perturbation of the clones in the algorithm (see section \ref{sec:rare_events_algorithms}).
Fig.~\ref{fig:pdf_drag_with_perturbation} shows that the statistics of the perturbed and unperturbed simulations are equivalent.

\bibliographystyle{jfm}
\bibliography{biblio}

\end{document}